\documentclass[AAPM,graphicx,nofootinbib]{revtex4-1}

\usepackage{color}
\usepackage[mathlines]{lineno}
\modulolinenumbers[5]
\usepackage{graphicx}
\usepackage{amsmath}

\usepackage{setspace}

\begin{document}


\title{Comprehensive procedure for personalized dosimetry in computed tomography} 



\author{S. Rosendahl}
\email[]{stephan.rosendahl@ptb.de}
\author{L. B\"uermann}
\email[]{ludwig.bueermann@ptb.de}

\affiliation{Physikalisch-Technische Bundesanstalt, Bundesallee 100, 38116 Braunschweig, Germany}

\author{M. Borowski}
\affiliation{St\"adtisches Klinikum Braunschweig gGmbH, Salzdahlumerstr. 90, 38126 Braunschweig, Germany}

\author{M. Kortesniemi}
\affiliation{HUS Medical Imaging Center, Helsinki University Central Hospital,  Haartmaninkatu 4, 00290 Helsinki, Finland}

\author{A. Kosunen}
\affiliation{STUK - Radiation and Nuclear Safety Authority, Laippatie 4, 00880 Helsinki, Finland}

\author{T. Siiskonen}
\affiliation{STUK - Radiation and Nuclear Safety Authority, Laippatie 4, 00880 Helsinki, Finland}

\author{V.-M. Sundell}
\affiliation{HUS Medical Imaging Center, Helsinki University Central Hospital,  Haartmaninkatu 4, 00290 Helsinki, Finland}

\date{\today}
\begin{abstract}
The purpose of this work is to develop viable procedures for verifying the applicability of personalized dosimetry in computed tomography (CT) using Monte Carlo-based simulations. Mobile equipment together with customized software was developed and used for rapid, non-invasive determination of equivalent source models of CT scanners under clinical conditions. Standard and anthropomorphic CT dose phantoms equipped with real-time CT dose probes at five representative positions were scanned. The accumulated dose  was  measured during the scan at the five positions. ImpactMC, a Monte Carlo-based CT dose software program, was used to simulate the scan. The necessary inputs were obtained from the scan parameters, from the equivalent source models and from the material-segmented CT images of the phantoms. Post-scan 3D dose distributions in the phantoms were simulated and the dose values calculated at the five positions inside the phantom were compared to measured dose values. 
Initial results were obtained by means of a General Electric Optima CT 660 and a Toshiba (Canon) Aquilion ONE. In general, the measured and calculated dose values were within relative uncertainties that had been estimated to be less than 10\,\%. The procedures developed, which allow the post-CT scan dose to be measured and calculated at five points inside anthropomorphic phantoms, were found to be viable and rapid. The procedures are applicable to any scanner type under clinical conditions. Results show that the procedures are well suited for verifying the applicability of personalized CT dosimetry based on post-scan Monte Carlo calculations. 
\hspace{3cm}
\end{abstract}


\maketitle 

\section{INTRODUCTION}
\label{sec:Introduction}

New hardware and software strategies in medical computed tomography (CT) have led to steady improvements in this type of imaging modality. Advanced X-ray tubes, new types of photon detectors for imaging, dynamic X-ray fluence as a function of the table position and the rotation angle (tube current modulation, TCM), spectral CT (e.g. dual source), spectral and spatial filtration, automatic selection of the tube potential, dynamic bow-tie (BT) filters and collimation are only some of the new hardware strategies \cite{Ginat2014, McCollough2006}. New software modules have been developed such as iterative reconstruction algorithms and, more recently, methods based on convolutional neural networks \cite{Yamashita2018}. All these new strategies have significant potential to lower the radiation dose to the patient while maintaining the necessary application-specific image quality. However, despite these promising developments, the patient dose in CT is still a concern. CT procedures deliver approximately 50-60\,\% of the collective effective dose from medical and dental exposures in many countries due to the relatively high dose of CT procedures compared with other diagnostic imaging modalities \cite{EC2014, NCRP2009}.

A recent topical review \cite{Kalender2014} provides an excellent overview about the past, current and future directions of X-ray CT dosimetry. Two measurable dose quantities must be indicated on every CT scanner: the volume computed tomography dose index $(CTDI_{\rm vol})$ and the dose-length product $(DLP)$ \cite{Shope1980, IEC2016}. $CTDI_{\rm vol}$ is defined for two cylindrical CTDI phantoms made of PMMA with diameters of 16 and 32\,cm and a typical length of 15\,cm, which are referred to as head and body phantoms. $CTDI_{\rm vol}$, indicated in units of mGy, can be interpreted as the mean dose deposited in a small cylindrical slice of the CTDI phantom. The $DLP$, indicated in units of mGy$\cdot$cm, is obtained by multiplying $CTDI_{\rm vol}$ by the total scan length and is thus correlated with the total dose of the scan. Both quantities are useful for quantifying X-ray tube dose output and comparing dose levels. Therefore, these two quantities are used as diagnostic reference levels \cite{ICRP2017}. Furthermore, they are used as dose indicators for acceptance and constancy tests. Although they are correlated with patient dose levels, they should not be regarded as a patient dose \cite{McCollough2011} because they are defined only for the two fixed-sized cylindrical PMMA phantoms and are not suitable for the variability of patient anatomy and size. In order to obtain generic stochastic radiation risk estimates, which are expressed in terms of an effective dose $(E)$, this quantity is evaluated from the $DLP$ by $E = k\cdot DLP$, where $k$ is a body-region-specific normalized effective dose conversion coefficient, usually calculated by means of MC simulations for reference scanners and reference patients. Such k-factors are tabulated in published reports, e.g. \cite{Bongartz2004, Shrimpton2006}. However, this method of dose estimation has many shortcomings, due not only to the known limitations of the basic dose quantities \cite{Boone2007, Dixon2010, Dixon2013, Dixon2014} if used in advanced-technology CT scanners (e.g. with TCM) or for stationary table CT but also to the fact that they are neither scanner-specific nor patient-specific.

In order to overcome the limitations given by the fixed sizes of the two CTDI phantoms, the {\it American Association of Physicists in Medicine} (AAPM) introduced the concept of {\it size-specific dose estimates} (SSDE) \cite{Boone2011, McCollough2014}. In the first step, a water equivalent diameter, $D_{\rm w}({\rm z})$, is determined at a longitudinal position z. $D_{\rm w}({\rm z})$ is defined as the diameter of a cylinder of water having the same average absorbed dose as the material contained in an axial plane of the scanned object. It is calculable for a material of any composition and quantifies the corresponding attenuation in terms of the attenuation in water. The SSDE at position z is obtained by $ SSDE{\rm (z)} = f(D_{\rm w}({\rm z})) \cdot CTDI_{\rm vol}$, where $f$ is a unit-less, empirically-derived factor that relates the radiation output of the scanner (which is quantified using $CTDI_{\rm vol}$) to the absorbed dose to soft tissue for a specific patient size or a specific phantom. $f(D_{\rm w}({\rm z}))$ is determined for a specific CTDI phantom size. The SSDE value of the whole scan is evaluated as the arithmetic average of $SSDE{\rm (z)}$. Although SSDE values correlate much better with the patient size and thus with the patient dose, they have the disadvantage that they are based on the conventional definition of CTDI with its known limitations. Furthermore, it is not possible to derive organ doses from SSDE values; thus, estimations of radiation risk are not possible.

\begin{figure}[ht]
 \includegraphics[width=500pt]{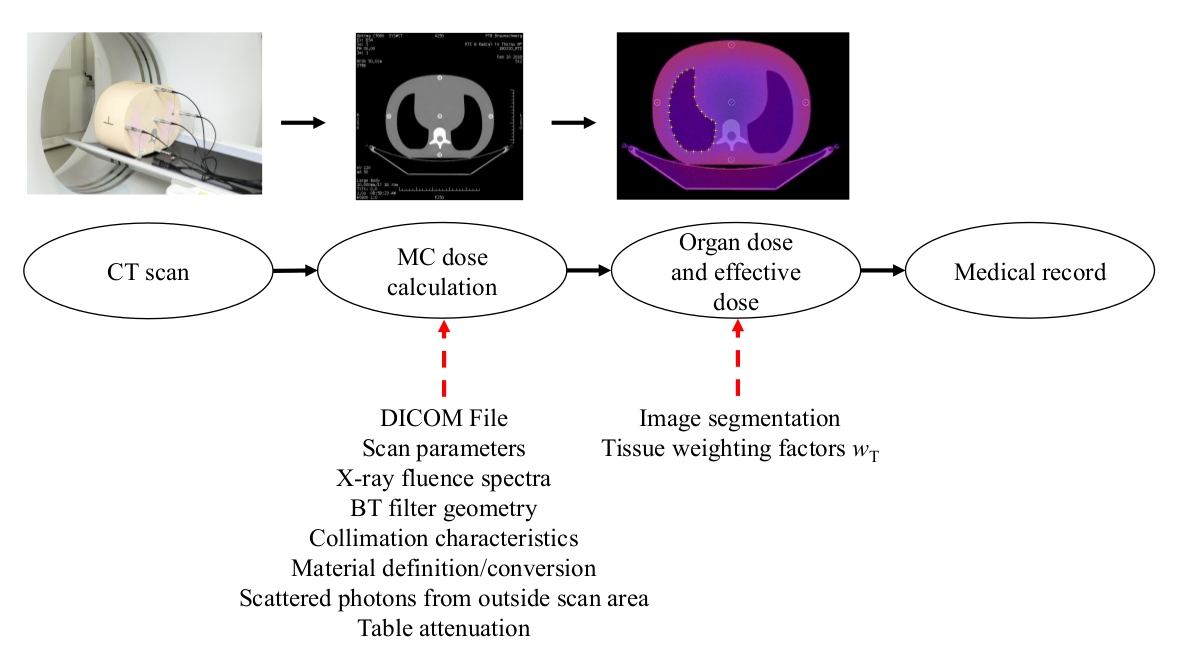}
 \caption{{\bf Procedure for personalized dosimetry in computed tomography.} A possible approach to personalized dosimetry starts from the CT scan, using the patient geometry as an input parameter. After scanner-specific and scan-specific parameters are inserted, the 3D dose distribution can be calculated. The procedure is validated with dose measurements in an anthropomorphic phantom, as indicated in the pictures.}
\label{fig:WorkflowPersDosimetry}
\end{figure}

A promising new approach of patient-specific dose estimates (PSDE), which was described in the above-mentioned review paper \cite{Kalender2014}, is based on fast Monte Carlo (MC) simulations and is independent of CTDI metrics. This approach makes use of the acquired patient CT data and combines it with a “best-fitting” pre-described voxel phantom to obtain a whole-body data set. Next, an MC dose calculation is performed by simulating the whole scan. An overlay of the calculated 3D dose distribution and the organ contours of the CT image allows organ dose values to be determined. The procedure, together with the necessary input information, is illustrated in Figure \ref{fig:WorkflowPersDosimetry}.

A fast MC dose calculation tool for use in clinical CT on-site and in real time has been developed and validated against measurements \cite{Schmidt2002, Deak2008, Myronakis2009, Chen2011}. Although this new approach has the potential to allow real-time scanner-specific and patient-specific dose estimations and indications on the scanner console to be made, it is still in the research stage. One reason for this may be the lack of standardized practical procedures for this application. Procedures for automatic segmentation of organs or anatomical structures in CT images are still challenging. Such segmentation can be performed manually but is a time-consuming process. However, new, promising procedures based on convolutional neural networks have been reported in literature \cite{Yamashita2018} and could represent a solution. Another problem is that the MC simulations require scanner-specific input data such as X-ray spectrum and subsequent filtration, including bow tie-shaped form filters, which are proprietary information that is not usually provided to the user. To overcome this restriction, some efforts have been made in recent years to develop methods that allow the determination of so-called ``equivalent X-ray source models'' that consist of a photon energy spectrum and filtration description that are based entirely on measured values \cite{Kruger2000, Turner2009, McKenney2014, Randazzo2015, Alikhani2016}. Finally, even if feasible procedures for PSDE are available, it may still be a time-consuming task to validate the dose simulation software by measurements at different scanner types.

The main purpose of this work is to develop standardized practical procedures for personalized CT dosimetry under clinical conditions based on the application of the above-mentioned PSDE approach. Except for the automatic segmentation of organ tissues in CT images, all other tools necessary for the realization and validation of the PSDE approach were (in principle) already available when this work was started. These tools are as follows: fast MC simulation codes for CT applications; non-invasive procedures for the determination of equivalent source models at different scanner types; and real-time dose detectors and anthropomorphic phantoms for the validation of calculated dose values by means of measurements. However, it was necessary to adjust and combine these tools to form a unique, rapid hardware and software package designed for the rapid application and verification of the PSDE approach. Mobile equipment and application-specific software were developed and used for rapid, non-invasive determination of equivalent source models of CT scanners under clinical conditions. These methods do not use the service mode of scanners. Standard CTDI and anthropomorphic CT dose phantoms were scanned, inside of which real-time CT dose probes were equipped at five representative positions. The accumulated dose at the five positions was measured during the scan. ImpactMC \cite{ABCT2014}, a fast Monte Carlo-based CT dose software program, was used to simulate the scan and to derive 3D dose distributions, as illustrated in Figure \ref{fig:WorkflowPersDosimetry}. The necessary inputs were obtained from the scan parameters, while the equivalent source models were determined with the mobile equipment and the material-segmented CT images of the phantoms. Post-scan 3D dose distributions in the phantoms were obtained, and dose values that had been calculated at the five detector positions inside the phantom were compared with the measurements. A comprehensive uncertainty analysis made it possible to assess the degree of consistency of measured and calculated values. The entire procedure was applied to two different scanner types, a GE Optima CT 660 and a Toshiba (Canon) Aquilion ONE. To our knowledge, this is the first comprehensive procedure that allows the complete PSDE approach on different scanners under clinical conditions to be verified.

In Section \ref{sec:Materials}, the CT scanners, the dose measurement systems and the mobile setup used to determine the equivalent source models are described. The ImpactMC software and its simulation and scan parameters is introduced as well as the polymethylmethacrylate (PMMA) and anthropomorphic phantoms and the related material conversion characteristics needed for the simulations. The equivalent source models for both CT scanners are given in Section \ref{sec:Results} followed by the comparison of simulated and measured dose distributions. The procedure for personalized dosimetry is demonstrated for a standard patient chest protocol applied to an anthropomorphic phantom using the GE Optima Optima CT 660. Finally, a discussion of the results is given in Section \ref{sec:Discussion}, followed by a summary in Section \ref{sec:Conclusion}.

\section{MATERIALS AND METHODS}
\label{sec:Materials}

\subsection{Description of CT scanners used for the study}
\label{sec:CTscanner}

For this study, two CTs manufactured by General Electric (GE) and Toshiba\footnote{In 2016, Canon Inc. took over the medical unit of Toshiba Corp.} were used. While the Optima CT 660 from GE is located at PTB for research purposes, the Toshiba Aquilion ONE is used for patient examination in clinical practice.

The GE Optima CT 660 (in the following referred to Optima 660) is the third version of GE's Optima family and is designed as a 64-slice multi-purpose volume CT for both day-to-day applications and advanced diagnostics \cite{GEHomepage}. It is used at PTB for investigating patient-specific CT dosimetry and characterizing image quality. Although it is employed only as a research CT, it has the same functions as other CTs used in clinical environments, including the standard patient protocols provided by GE. The machine allows for tube voltages of 80\,kV, 100\,kV, 120\,kV and 140\,kV with a maximum collimation of 40\,mm and is equipped with one large BT filter and one small BT filter.

The Toshiba Aquilion ONE used at the St\"adtisches Klinikum hospital in Braunschweig is a 3rd generation CT with 320 slices. With a total collimation of 16\,cm, large anatomic areas can be scanned within one rotation. Hence, it has been claimed to be the first dynamic volume CT scanner on the market \cite{Canon}. Tube voltages of 80\,kV, 100\,kV, 120\,kV and 135\,kV can be selected. Furthermore, it is equipped with three different BT filters: small, medium and large \cite{ImPACT2009}. 

\subsection{Devices for dose measurements}
\label{sec:Detectors}

The dose measurements are performed with various detectors. Different ionization chambers are used: a farmer type chamber (Radcal, type RC0.6) and two pencil-type CT chambers 100\,mm and 300\,mm in length (PTW, type 30009 and 30017, respectively). While the RC0.6 is capable for measurements of the air kerma $K_{\rm a}$, the CT chambers are used to measure the air kerma length product $K_{\rm a} L$. The current from the ionization chambers is measured with an electrometer (PTW, type UNIDOS). $K_{\rm a}$ and $K_{\rm a} L$ are given by the following expressions:

\begin{align}
K_{\rm a}=M_{{\rm Q_0}}\cdot N_{K_{\rm a, Q_0}}\cdot k_{{\rm Q, Q_0}} \cdot k_{\rho} \\
K_{\rm a} L=M_{{\rm Q_0}}\cdot N_{K_{\rm a, Q_0}L}\cdot k_{{\rm Q, Q_0}} \cdot k_{\rho}.
\label{eq:Detectorcalibration}
\end{align} 

Here, $M_{{\rm Q_0}}$ is the charge measured at the electrometer and $N_{K_{\rm a, Q_0}}$ and $N_{K_{\rm a, Q_0}L}$ are the calibration factors at reference quality ${\rm Q_0}$ determined using the primary air kerma standard at the X-ray facilities of PTB. $k_{{\rm Q, Q_0}}$ is the quality correction factor that takes into account the energy dependence of the detector for different X-ray qualities that can appear in the CT, e.g. 80\,kV against 140\,kV. This value is set to $k_{{\rm Q, Q_0}}=1$ since the response of pencil-type ionization chambers with respect to the air kerma is nearly independent from the spectra in the energy range between 30\,keV and 150\,keV \cite{Siiskonen2018}. Small deviations are taken into account via the uncertainties given in Table \ref{tab:UncertaintyBudget_GE}, which has been validated using the methods presented in \cite{Siiskonen2018}. Uncertainty estimations within this work were performed according to the {\it Guide to the expression of uncertainty in measurement (GUM)} \cite{BIPM2008} using the GUM Workbench \cite{Metrodata2017}.

\begin{table}[b]
 \caption{{\bf Uncertainty budget for air kerma measurements on CT scanners.} This table summarizes the uncertainty budget for dose measurements on CT scanners presented in this work using different ionization chambers and semiconductor detectors.}
\vspace{10pt}
 \begin{tabular}{lccccccccccccc}
 Description of & \vline &Symbol&\vline& Uncertainty& \vline & \multicolumn{3}{c}{Pencil type chamber}& \vline & Farmer type chamber & \vline & Semiconductor  \\
 component& \vline & &\vline&  type & \vline & 10\,cm  & \vline & 30\,cm & \vline& RC0.6 & \vline & dose profiler \\
  & \vline & &\vline&    & \vline & $u_{{\rm A,B}}$ (\,\%)  & \vline & $u_{{\rm A,B}}$ (\,\%) & \vline & $u_{{\rm A,B}}$ (\,\%) & \vline &$u_{{\rm A,B}}$ (\,\%)\\

\hline
\hline
Repetition (example)&           \vline & 		$\sigma_{{\rm rep}}$						&\vline& A & \vline & 	0.46 & \vline & 0.28 & \vline & 0.30	 & \vline & 1.13  \\
\hline
 
Calibration factor& 	\vline & $N_{K_{{\rm a, Q_0}}}$ &\vline&  B & \vline &  0.40  & \vline & 0.20 & \vline&  0.40 & \vline & - \\
(RQT-9)& 							\vline &  								&\vline&    & \vline &   	& \vline & & \vline&  & \vline & \\ 
Energy dependence & 	\vline & $k_{{\rm Q, Q_0}}$&\vline&  B & \vline & 0.45  & \vline & 0.81 & \vline& 0.38 & \vline & (0.50) \\
(Free in air)& 				\vline &  							&\vline&    & \vline &   & \vline&   & \vline &  & \vline & \\ 
Energy dependence & 	\vline & $k_{{\rm Ph}}$&\vline&  B & \vline & -  & \vline &  -& \vline& - & \vline & 1.05 \\
(In phantom)& 				\vline &  							&\vline&    & \vline &   & \vline &   & \vline&  & \vline & \\ 
Reproducibility &			\vline & 						 		&\vline&  B & \vline & 	0.80 & \vline & 0.43 & \vline& 0.34 & \vline & 0.59 \\
(e.g. positioning)& 	\vline &  							&\vline&    & \vline &   & \vline &  0.10 & \vline&  & \vline & \\ 
Air density correction&					\vline & $k_{\rho}$ 		&\vline&  B & \vline & 0.14	 & \vline& 0.14 & \vline & 0.14 & \vline & - \\
\hline
Combined type B& \vline & 	$\sigma_{\rm B}$	&\vline&    & \vline &  1.01 & \vline & 1.01 & \vline& 0.66 & \vline &  1.21\\ 
uncertainty  $(k=1)$ & \vline & 		&\vline&   & \vline &   & \vline &  & \vline&  & \vline &  \\ 
\hline
Combined uncertainty& \vline & 	$\sigma_{\rm c}$		&\vline&   & \vline & 1.11  & \vline & 1.03 & \vline& 0.73 & \vline & 1.65 \\ 
 $\sqrt{u_{\rm A}^2+u_{\rm B}^2}$  $(k=1) $& \vline & 	&\vline&   & \vline &    & \vline &   & \vline&   & \vline &   \\ 
 \end{tabular}
 \label{tab:UncertaintyBudget_GE}
 \end{table}

Finally, $k_{\rho}$ is the correction factor for the air density (see Table \ref{tab:UncertaintyBudget_GE}) calculated from ambient pressure and temperature, which is measured using quality-controlled, high accuracy sensors. 

The RC0.6 is also used in combination with a digitizer read-out (Radcal, ACCU-Gold) that allows time-resolved measurement of the dose rate to be taken.

In addition to ionization chambers, semiconductor detectors (RTI, dose profiler) are used for measurements of the air kerma. The advantage of these detectors is their high spatial resolution of less than 0.25\,mm \cite{RTI2015} in longitudinal (z-) direction and their higher sensitivity. Furthermore, the measurement system (RTI, Barracuda) allows five sensors be be connected for time-saving simultaneous measurements of $K_{\rm a}$, which is given by:

 \begin{equation}
K_{\rm a}=M_{{\rm Q_0}}\cdot N_{K_{{\rm a, Q_0}}}\cdot k_{{\rm Q, Q_0}} \cdot k_{{\rm Ph}}.
\label{eq:dose profilercalibration}
\end{equation} 

In contrast to ionization chambers, air density correction is not needed; here, the correction of the energy dependence is more complicated. The Barracuda system already gives a calibrated value for air kerma $K_{\rm a}=M_{{\rm Q_0}}\cdot N_{K_{{\rm a, Q_0}}}$, but the manufacturer uses only one X-ray quality (RQR-9) for calibration. For tube voltages between 80\,kV and 140\,kV, significant differences of up to $\approx$\,30\,\% can be observed due to the spectral dependence of the detector. Hence, additional calibrations with qualities from the ISO4037 narrow-spectrum series \cite{ISO-N} are performed at PTB in order to calculate correction factors $\rm{k_{{\rm Q, Q_0}}}$ for each tube potential using the CT-specific spectra obtained from Al attenuation measurements (compare \cite{Siiskonen2018}). The comparison of a spectrally corrected dose profiler with the RC0.6 chamber, both of which measure the air kerma successively at the iso center of the Optima 660, leads to differences of less than 2\,\%. This is in agreement with the uncertainties stated in Table \ref{tab:UncertaintyBudget_GE} and indicates that the correction factors are correct.  However, the spectrum of the CT changes inside the phantom due to attenuation and scattering. For this reason, an additional factor $k_{{\rm Ph}}$ is introduced, taking the spectral modulation caused by the material of the phantom into account. $k_{{\rm Ph}}$ is estimated by comparing the air kerma measurement of the dose profiler after spectral correction using $k_{{\rm Q, Q_0}}$ with the values measured with the ionization chamber RC0.6 in the phantom. The direct calibration is performed for the different standard and anthropomorphic phantoms using the tube voltages that are relevant for this study. For the usage of ionization chambers inside the phantom, $k_{{\rm Ph}}$ is not needed, since the energy dependence is negligible and covered by the uncertainty, as mentioned above.

\subsection{Mobile measurement equipment and equivalent CT source determination}
\label{sec:MobileSetup}

\begin{figure}[htb] 
    \includegraphics[width=290pt]{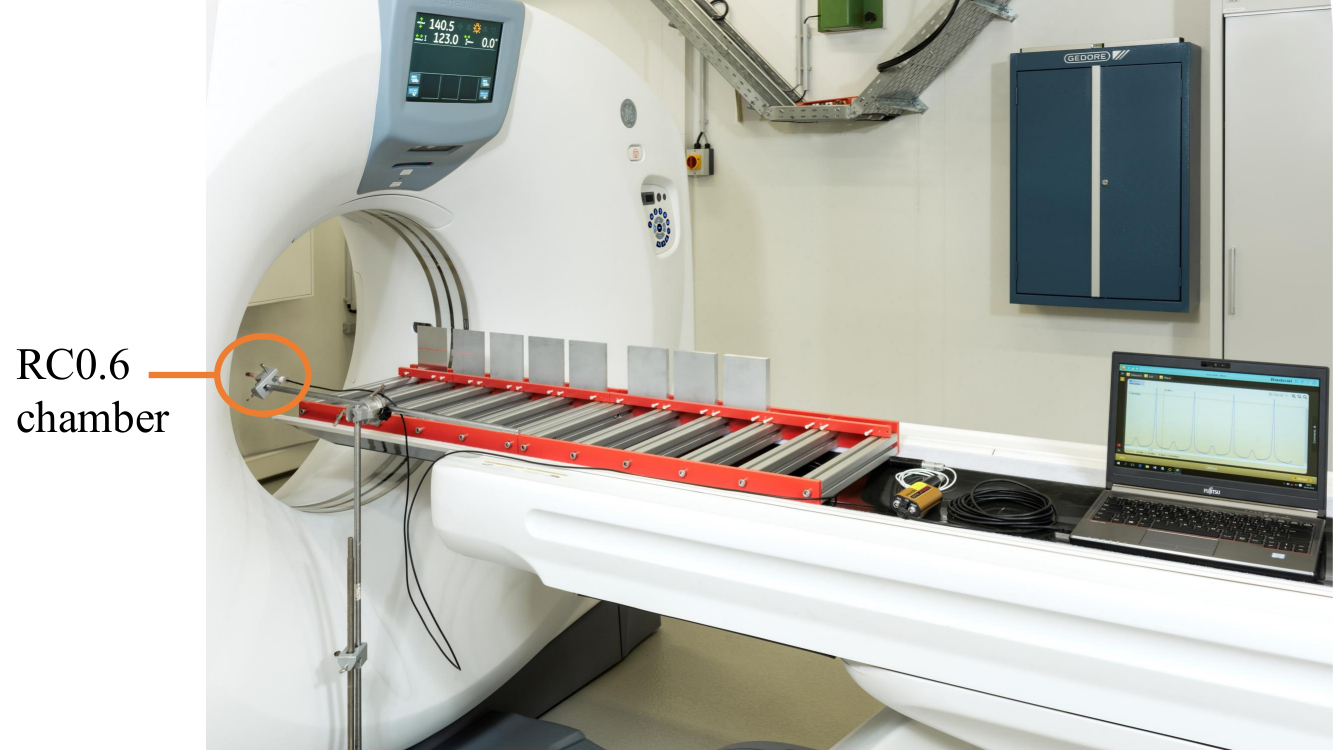} 
\caption{\textbf{Mobile measurement setup.} With this setup, aluminum attenuation curves and form filter characteristics can be investigated that allow the computation of equivalent CT source models. Aluminum sheets of increasing thickness are placed on the patient table, while the dose rate is measured with an ionization chamber and time-resolved readout.} 
\label{fig:MobileSetup}
\end{figure}

X-ray spectra can be determined by measuring aluminum attenuation characteristics. This method is well established and widely used \cite{Kruger2000, Turner2009, Alikhani2016, McKenney2014, Randazzo2015}. Conventional spectrometry using semiconductor detectors made of materials such as germanium is complex and currently only possible by measuring Compton-scattered photons to obtain the original spectrum via deconvolution \cite{Matscheko1988, Bazalova2007, Duisterwinkel2015}. Measuring X-ray attenuation as a function of increasing aluminum thickness with an ionization chamber provides easy access to a very accurate approximation of the real spectrum. Typical experimental procedures using the service mode of the CT scanner are presented in \cite{Turner2009, Alikhani2016}, while more advanced setups have been developed using rotating X-ray sources and real-time dose measurements \cite{McKenney2014, Randazzo2015}. For the mobile measurement system, a time-resolved read-out is used in order to measure (in spiral mode) the attenuation caused by aluminum plates of different thickness that are placed on a holding frame on the patient table, as shown in Figure \ref{fig:MobileSetup}. For this purpose, the RC0.6 probe together with the ACCU Gold digitizer is fast enough for rotation times of $\approx 1$\,s.

The detector is placed on the x-axis away from the iso center to match $z=y=0$ with a distance between the source and the iso center of $x=\rm{r_{\rm P}\approx 25\,cm}$. Hence, contamination of the signal by scattered photons from the aluminum sheets on the one hand and back-scattered photons from the detector array of the CT on the other hand are reduced. The collimation in z-direction $\rm{\Delta_Z}$ should be chosen in such a way that the ionization chamber is fully covered by the beam, e.g., $\rm{\Delta_Z \approx 2\,cm}$. A larger collimation produces additional scattered photons that may influence the results. The measurement procedure includes an initial overview scan in order to define scan areas centered at the single aluminum plates. This simplifies the data analysis and ensures that a certain distance between the scan area and the holding frame is maintained in order to reduce scattering from the frame. Typically, nine scan areas are defined: one free in air without aluminum as a reference and eight aluminum sheets with thicknesses between 1\,mm and 20\,mm and a purity of 99.999\,\%. The homogeneity in the thickness of the plates, as well as their absolute thicknesses, have been measured on site and shown to be better than 1\,\%. The parameters normally chosen for the scan are a rotation time of 1\,s  and a pitch of $\approx 0.5  $. The pitch depends on the table velocity and collimation and the possible values are scanner-specific. Hence, the total scan time varies for the different manufacturers, but is in the range of $\approx 2$\,min.

\begin{figure}[ht]
   \includegraphics[width=300pt]{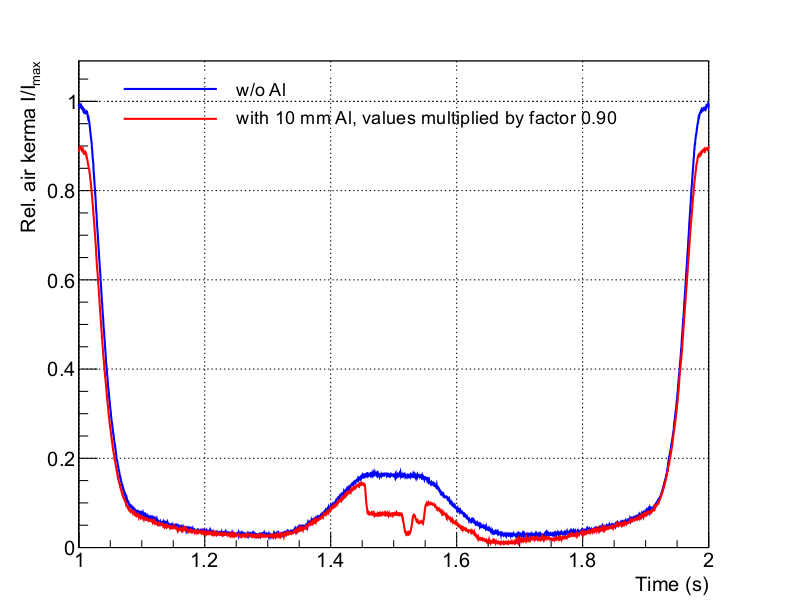} 
\caption{\textbf{Time-resolved dose measurement of aluminum attenuation characteristics.} With experimental geometry, the time-resolved dose curve shows characteristic behavior caused by the inverse square law and attenuation of the BT filter. The region of interest where the aluminum sheet, the detector and the source are aligned shows the attenuation caused by a 10\,mm layer of aluminum. The attenuated dose values with aluminum are multiplied by a factor of 0.9 in order to make both curves distinguishable for the reader.} 
\label{fig:HVL_Curve1}
\end{figure}

In Figure \ref{fig:HVL_Curve1}, the time-resolved dose rate for one rotation of the X-ray source is shown. The characteristic behavior is caused by the inverse square law and the attenuation of the BT Filter. The maximum of intensity at 1\,s and 2\,s are the position where the source and the detector are at their closest position.  The region of interest where the aluminum sheet, the detector and the source are aligned (at $\sim 1.5\,s$) shows the attenuation and is gained after a half rotation when the source, the aluminum sheet and the detector are aligned (compare Figure \ref{fig:MobileSetup}). It is expressed as the ratio $\kappa_{{\rm Meas}}$ of $K_{\rm x}$ to $K_{\rm 0}$, which are the air kerma with and without the added aluminum sheet of the thickness $x_{{\rm Al}}$:
\begin{equation}
 \kappa_{{\rm Meas}} = \frac{K_{\rm x}}{K_{\rm 0} }
\label{eq:K_Meas}
\end{equation}

The signal inhomogeneities after the attenuation from the Al layer are caused by the patient table and the holding frame. In a further analysis, values of $\kappa_{{\rm Calc}}$ are calculated by:

\begin{equation}
 \kappa_{{\rm Calc}}=\frac{K_{\rm x}^{{\rm Calc}}( x_{{\rm Al}})}{K_{\rm 0}^{{\rm Calc}}}=\frac{\int \mathit{\Phi}_E e^{-\mu_{{\rm Al}}(E)\cdot d_{{\rm Al}}} e^{-\mu_{{\rm Al}}(E)\cdot x_{{\rm Al}}} E(\frac{\mu_{{\rm en}}}{\rho})_{{\rm Air}} dE}{\int \mathit{\Phi}_E e^{-\mu_{{\rm Al}}(E)\cdot d_{{\rm Al}}} E(\frac{\mu_{{\rm en}}}{\rho})_{{\rm Air}} dE},
\label{eq:K_Calc}
\end{equation}

where $ \mathit{\mathit{\Phi}}_{\rm E}$ is the non-filtered X-ray spectrum and $d_{{\rm Al}}$ is the quality equivalent filtration thickness of aluminum. Note that $d_{{\rm Al}}$ represents the total filtration composed of the inherent filtration of the tube and the additional filtration from the center of the BT filter and from the Mylar window. 
$\mathit{\Phi}_{\rm E}$ is calculated using the SpecCalc software tool \cite{Poludniowski2007a, Poludniowski2007b, Poludniowski2009}, which requires the tungsten anode angle and the tube high voltage as inputs. $d_{{\rm Al}}$ is obtained by minimizing $\Delta(d_{{\rm Al}})$, given as

\begin{equation}
\Delta(d_{{\rm Al}})=\sum_{{\rm i}}\left| \kappa_{{\rm Meas}}(x_{\rm i})-\kappa_{{\rm Calc}}(d_{{\rm Al}}, x_{\rm i})\right|.
\label{eq:Delta_dAL}
\end{equation}

In-house analysis software reads the measured data and identifies the regions of interest for the calculation of the attenuation curve. Due to the small pitch factor and collimation, multiple rotations per Al sheet can be used to calculate the average $\kappa_{{\rm Meas}}(x_{\rm i})$.

\begin{figure} 
   \includegraphics[width=240pt]{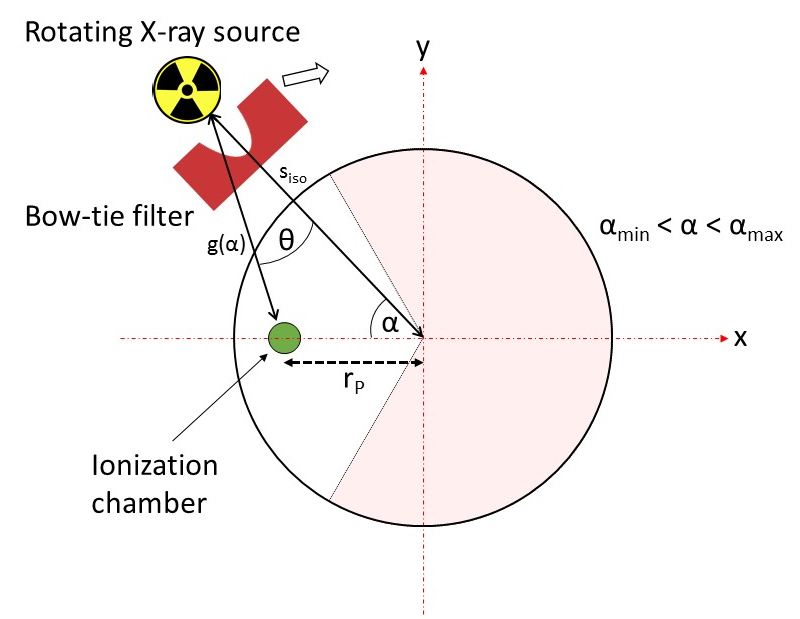} 
	\quad
    \includegraphics[width=240pt]{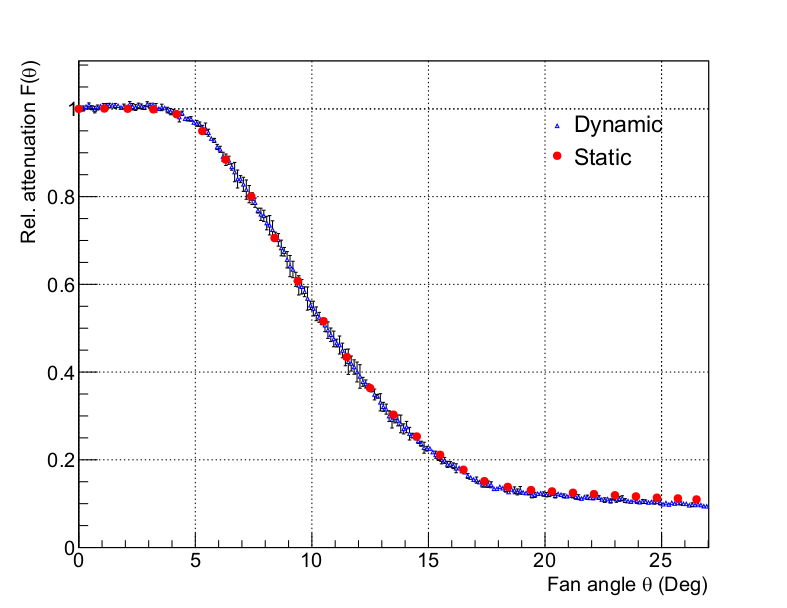} 
\caption{\textbf{Bow tie filter characterization using the COBRA formalism.} Schematic overview of the COBRA geometry (left) and the relative attenuation profile of the large BT of the Optima 660 (right). With the detector placed away from the iso center, the relative attenuation of the BT can be measured with a rotating X-ray source as the function of the rotation angle $\alpha$ using a time resolved readout. The distance between the detector and the iso center $r_{\rm P}$ restricts the maximum possible fan angle $\theta$ that can be measured and defines the rotation angles for the analysis between $\alpha_{{\rm min}}$ and $\alpha_{{\rm max}}$ in the area marked red.}
\label{fig:Cobra_Scheme}
\end{figure}

The BT filter is characterized using the COBRA formalism ({\bf c}haracterization {\bf o}f {\bf b}ow-tie {\bf r}elative {\bf a}ttenuation) proposed by Boone in 2010 \cite{Boone2010}. The geometry for the mobile measurement system has been chosen in order to additionally collect the data relevant for the determination of the attenuation profile without additional hardware re-arrangements, which represents a significant advantage of the system. The attenuation is measured performing a dedicated free-in-air scan with a collimation of $\rm{\Delta_Z = 4\,cm}$. A smaller collimation is not possible because it prevents full coverage of the detector when the distance between the source and the detector is too small. For the following COBRA analysis, which was performed to determine the attenuation characteristic of the BT filter, we strictly followed the calculation process of Boone \cite{Boone2010}, transforming the measured dose from the time domain $ I{\rm(t)}$ into the rotation domain $ I(\alpha)$, with $\alpha$ being the rotation angle, as shown in Figure \ref{fig:Cobra_Scheme}.

It is assumed that, at rotation angles $\alpha=0$ and $\alpha=\pi$, where the X-ray source, the probe and the iso center are aligned (collinear), the thickness of the BT filter is zero for the calculations, as the attenuation in the center of the BT (fan angle $\theta=0$) is at its minimum \cite{Boone2010, McKenney2011}. The detector measures the highest possible intensity $I_{\rm 0}{\rm(0)}$ at $\alpha=0$, which modulates according to the inverse square law while the source rotates. $ I_{\rm 0}(\alpha)$ is calculated using the distances between the source and the iso center $ s_{{\rm iso}}$ and between the source and the detector $ g(\alpha)$. Furthermore, $ I_{\rm 0}(\alpha)$ used the measured intensity $I_{{\rm att}}(\alpha)$ as reference to determine the attenuation ${F(\theta)}$ caused by the BT  \cite{Boone2010, McKenney2011}:

\begin{equation}
F(\theta)=\frac{I_{{\rm att}}(\alpha)}{I_{\rm 0}(\alpha)} \qquad with \qquad I_{\rm 0}(\alpha) = I_{\rm 0}{\rm (0)}\cdot \left(\frac{s_{{\rm iso}}-r_{\rm P}}{s_{{\rm iso}}}\right)^2\cdot \left(\frac{s_{{\rm iso}}}{g(\alpha)}\right)^2.
\label{eq:F_theta}
\end{equation}

In Figure \ref{fig:Cobra_Scheme} (right), the relative attenuation profile measured on an Optima 660 is shown for the large BT. In order to perform a quality check of the implementation, the attenuation curve obtained from the dynamic COBRA mode (blue dots) is compared with the curve obtained from static step-and-shot measurements (red dots), which is in good agreement. The details of the static method using a fixed X-ray source in the service mode of the CT scanner can be found elsewhere \cite{Alikhani2016}. The analysis was performed using rotation angles in the range of ${\rm \alpha_{min} < \alpha <  \alpha_{max}}$, while the minimum and maximum rotation angles ${\rm \alpha_{min}}$ and ${\rm \alpha_{max}}$ depend on $ r_{\rm P}$. The measurements from the other angles are neglected in this work, since the distance between the probe and the detector is very small; this leads to significant contamination, disturbing the attenuation profile, which should be measured in narrow beam geometry. 

In the standard COBRA formalism, the thickness at the center of the BT is assumed to be zero for the calculations. This initial thickness at the center is usually included in the equivalent filtration of the spectrum that is achieved in dynamic measurements of the aluminum attenuation characteristics. Consequently, in the MC simulations, the equivalent BT filter can be used only in combination with the associated X-ray spectrum containing the correct equivalent filtration, which is a potential source of errors when incorrect files are combined. To reduce this uncertainty, the measured total filtration $ d_{{\rm Al}}$ is treated as part of the BT filter, allowing the unfiltered spectrum to be used for the simulation. Here, the advantage is that the Al-equivalent filter geometry derived in this way contains all the specific information from the bow tie as well as from the total filtration. The unfiltered X-ray spectrum used in the simulations is independent from the BT filter used.  

From the relative attenuation $ F(\theta)$, an equivalent Al filter is calculated as described in e.g. \cite{Alikhani2016, Boone2010}. 

\subsection{MC simulation with ImpactMC using homogenous PMMA and anthropomorphic phantoms}
\label{sec:ImpactMC}
ImpactMC is a simulation tool that traces its origins to works by Schmidt and Kalender \cite{Schmidt2002}. For this study, Version 1.5.1 of ImpactMC in GPU mode is used. The user can control a wide range of parameters that allow scanner-specific X-ray spectra and BT-filter geometries to be included. Furthermore, asymmetric collimation and tube current modulation can be incorporated \cite{ABCT2014}, although these techniques are not investigated here. Of crucial importance for obtaining accurate simulation results is the material conversion feature of ImpactMC, which is needed in order to link the Houndsfield units (HU) from the DICOM file to the specific material. 

The validation process presented in this work was carried out with two different phantoms: the standard CTDI phantoms made of PMMA with a diameter of 32\,cm (body) and a ``Thorax-Medium Adult'' anthropomorphic phantom (Mod. Nr. 007TE-17) manufactured by the CIRS company. Details of the physical dimensions can be found online \cite{CIRS2013}. Both the homogenous and the anthropomorphic phantom are successively equipped with the 100\,mm pencil chamber to measure conventional CTDI values and with five dose profilers for the simultaneous measurement of point doses. 

Since both PMMA and anthropomorphic phantoms are used in this study, the material conversion file contains values for PMMA as well as for lung tissue, bone and water. The ranges for material conversion used in this work are given in Table \ref{tab:Materialconversion}. The water equivalent material of the phantom represents, to some extent, the soft tissue. Hence, the range for the Houndsfield units of water is enlarged. In addition, the ImpactMC software requires the consecutive definitions of the material conversion; changing the minimum HU for water to a value of e.g. -30 would enlarge the range for lung tissue. Furthermore, materials related to the patient table such as carbon-fiber and aluminum are included in the conversion file as well. Since the composition of the carbon fiber is unknown, an equivalent material has been defined that is made from carbon ($\approx 92$\,\,\% mass fraction), hydrogen ($\approx 5$\,\,\%) and lead ($\approx 3$\,\,\%). The fraction of lead has been calculated using the attenuation of the table (as measured with an ionization chamber), the geometry of the table from the DICOM file, the CT spectra and the mass attenuation coefficients of the materials mentioned above. In summary, an equivalent table material was calculated that mimics the attenuation properties of the table in order to achieve accurate simulation results.

 \begin{table}[htb]
 \caption{{\bf Material conversion from Houndsfield units.} Material classification is necessary for accurate simulation results due to the different dependencies on the photon energy and the atomic number of the photon interaction cross section of the different elements.}
\vspace{10pt}
 \begin{tabular}{ccccc}
  Material& \vline & HU min & \vline & HU max \\
\hline
Air & \vline & - & \vline & -850 \\
Lung & \vline & -850 & \vline & -750 \\
H$_2$O (soft tissue)& \vline & -750 & \vline & 50 \\
PMMA & \vline & 50 & \vline & 190 \\
Carbonfiber & \vline & 190 & \vline & 600 \\
Bone & \vline & 600 & \vline & 900 \\
Aluminum & \vline & 900 & \vline & 10000 \\
 \end{tabular}
 \label{tab:Materialconversion}
 \end{table}

In order to normalize the simulated values in absolute values of air kerma, the scanner output is used. To this end, the air kerma is measured free-in-air at the iso center of the scanner using the RC0.6 probe mounted at the front of the aluminum frame. A large collimation ($ \approx 40$\,mm) ensures the full coverage of the detector and gives the peak air kerma of the beam $K_{\rm a}^{\rm P}$. This value is used in ImpactMC for normalization and is inserted into the ``AirKerma [mGy/100\,mAs]'' field. Unfortunately, taking the average air kerma alone and assuming a rectangular beam profile with a nominal collimation are not sufficient to achieve correct results. Doing so would lead to an underestimation of the dose in the simulation, since Heel and penumbra effects as well as scattered radiation are not included. These effects deform the assumed rectangular shape; in particular, the penumbra effect, which is caused by a non-perfect collimation, enhances the deposited dose and needs to be taken into account. 

Hence, additional measurements were made using a 100\,mm pencil-type chamber to measure the free-in-air CTDI  $CTDI_{{\rm free\, air}}$ for different collimations of interest in the iso center. $CTDI_{{\rm free\, air}}$ is defined as the air kerma length product measured free-in air at the iso center of the CT and normalized to the nominal collimation. It includes the different effects, mentioned above, as well as scattered radiation from the BT filter. In order to realize the proper normalization from measuring  $CTDI_{{\rm free\, air}}$, the property labeled as ``total beam collimation'' in ImpactMC is set. For our purposes, the total beam collimation $ z_{{\rm Tot}}$ is calculated from the nominal beam width $z_{\rm N}$ using the ratio of $CTDI_{{\rm free\, air}}$ and peak air kerma $K_{\rm a}^{\rm P}$:

\begin{equation}
z_{{\rm Tot}}=z_{\rm N}\cdot \frac{CTDI_{{\rm free\, air}}}{K_{\rm a}^{\rm P}}=\frac{K_{\rm a} L}{ K_{\rm a}^{\rm P}}.
\label{eq:ztot}
\end{equation} 

The total beam width is typically larger than the nominal beam width. For example, using ${\rm{z_N}=10\,mm}$ and measured values for the peak dose and the $CTDI_{{\rm free\, air}}$ at 120\,kV of $K_{{\rm a}}^{{\rm P}}={\rm 21.30\,mGy/100\,mAs}$ and $CTDI_{{\rm free\, air}}={\rm 28.26\, mGy/100\,mAs}$ leads to $z_{{\rm Tot}}=13.3$\,mm (compare Table \ref{tab:K0}).

For broader collimations (e.g. 160\,mm), a longer, commercially available pencil-type chamber (300\,mm) was used.

\begin{table}[ht]
 \caption{{\bf Uncertainty budget for simulations of 3D dose distributions with ImpactMC.} The uncertainty budget for the 3D dose simulations from CT images using the ImpactMC software is estimated. Different sources of uncertainties are evaluated and the results are collected.}
\vspace{10pt}
 \begin{tabular}{lccccccc}
 Description of & \vline &Symbol&\vline& Uncertainty& \vline & Effect on simulation  \\
 component& \vline & &\vline&  type & \vline & with Impact MC   \\
  & \vline & &\vline&    & \vline & $u_{{\rm A,B}}$ (\,\%)  \\

\hline
\hline
Starting angle - axial &           \vline & 		$\sigma_{\alpha}$						&\vline& B & \vline & 	0.9  \\
   - helical&           \vline & 						&\vline&  & \vline &  7.5  \\
Repetition &           \vline & 		$\sigma_{{\rm Stat}}$						&\vline& B & \vline & 	0.8 &   \\
(Statistic uncertainty) &           \vline & 						&\vline&  & \vline &  &   \\
Total beam collimation - point&           \vline &$\sigma_{{\rm BC}}$ &\vline& B & \vline & 	0.3 &   \\
 (- air kerma length product $K_{\rm a} L$)&           \vline & 							&\vline&  & \vline & 	(1.2)&   \\
Air kerma normalization&           \vline & 		$\sigma_{\rm N}$						&\vline& B & \vline & 	0.7 &   \\

Shape of BT&           \vline & 		$\sigma_{{\rm BT}}$						&\vline& B & \vline & 	2.5 &   \\

\hline
Combined uncertainty& \vline & 			&\vline&   & \vline &   \\ 
 $\sqrt{u_{\rm B}^2}$ ,  $(k=1) $& \vline & &\vline&   & \vline &    \\
Axial ($K_{\rm a} L$)& \vline & $\sigma_{\rm c}$ 	&\vline&  & \vline &    2.9 (3.1)  \\
 Helical ($K_{\rm a} L$)& \vline & 	&\vline&   & \vline &    8.0 (8.1)  \\ 
 
 \end{tabular}
 \label{tab:UncertaintyBudget_ImpactMC}
 \end{table}

The uncertainty of the dose simulation is estimated by varying single input parameters while keeping the rest of the parameters fixed. In the first attempt, it is assumed that the different parameters have negligible correlations. The summary of these studies is presented in Table \ref{tab:UncertaintyBudget_ImpactMC}. The statistical uncertainty $\sigma_{{\rm Stat}}$ is defined by the number of photons ($4\cdot 10^9$) and the number of voxels, which is $512\times 512\times 201$ (voxel dimensions $0.98\times 0.98\times 2.5\,{\rm mm}^3$) for all geometries and was estimated by repeating the same simulation multiple times at $\sigma_{{\rm Stat}}=0.8\,\%$ for $k=1$. Other effects such as the uncertainty in the total collimation $z_{\rm{Tot}}$ and the normalization related to the air kerma using $K_{\rm a}^{\rm P}$ lead to similar values below 1\,\%. Except for the uncertainty in the simulation of the air kerma length product, the uncertainty from $z_{{\rm Tot}}$ is slightly higher, which is expected due to the penumbra effects. The construction of the equivalent BT filter leads to a much higher contribution to the final uncertainty. From the multiple measurements performed on the Optima 660, the filters with minimum and maximum thickness are extracted and compared to the dose results performed with the average filter. The resulting uncertainty is calculated to be $2.5\,\%$, which is a conservative estimation. The major contributor in this example is related to the rotation angle $\alpha$ of the X-ray tube. The starting angle of the exposure is an important input parameter, especially for helical scans with multiple rotations. Since the information of the angle is not given by the CT console nor by the header of the DICOM file, its uncertainty has been estimated by performing multiple simulations with different starting angles between $0^\circ$ and $270^\circ$ and averaging the results. The standard deviation was calculated to be $0.9\,\%$ for the axial mode and for one rotation in the first measurements. However, the standard deviation increases drastically to $7.5\,\%$ when helical scan modes and multiple rotations are used. For the latter case, the combined uncertainty $\sigma_{\rm c}$ is calculated by means of a quadratic summation to $\sigma_{\rm c}=8.0\,\%$ for point dose simulations and $8.1\,\%$ for simulated values of the air kerma length product $(k=1)$. For the axial mode, the combined uncertainty is calculated to be $\sigma_{\rm c}=2.9\,\%$ for point dose simulations and $3.1\,\%$ $(k=1)$ for simulated values of the air kerma length product.

\section{RESULTS}
\label{sec:Results}
\subsection{Determination of equivalent source models}
\label{sec:GEOptima}

\subsubsection{GE Optima CT 660}

The average Al thicknesses $d_{{\rm Al}}$ from the determination of the total filtration, as described in Chapter \ref{sec:Materials}, are given in Table \ref{tab:InhFiltrationValues}, organized according to the different BT filters and tube voltages. In total, eight measurements have been taken to date for the Optima 660 using the mobile measurement setup. The standard deviation for the small filter varies between 1.3\,\% and 2.7\,\%, while the corresponding values for the large filter are slightly higher, ranging between 2.1\,\% and 3.8\,\%. 

\begin{table}[ht]
 \caption{{\bf Total aluminum filtration of the Optima 660.} The values for the total filtration $d_{{\rm Al}}$ are given for the different tube voltages and BT filters and are average values from eight single determinations using the mobile setup.}
\vspace{10pt}
 \begin{tabular}{ccccccccccccc}
  Voltage & \vline &\multicolumn{5}{c}{Small filter}   & \vline &\multicolumn{5}{c}{Large filter} \\
	& \vline & $d_{{\rm Al}}$ [mm] & \vline & St. dev. [\,\%] & \vline & $1\sigma_{\chi}$ [\,\%] & \vline & $d_{{\rm Al}}$ [mm] & \vline & St. dev. [\,\%] & \vline & $1\sigma_{\chi}$ [\,\%] \\
\hline
\hline
80\,kV & \vline & {\bf 7.464} & \vline & 2.7 & \vline & 2.6 & \vline & {\bf 10.30} & \vline & 3.8 & \vline & 2.8\\
100\,kV & \vline & {\bf 7.376} & \vline & 2.2 & \vline & 1.9 & \vline & {\bf 10.26} & \vline & 2.1 & \vline & 2.2\\
120\,kV & \vline & {\bf 7.398} & \vline & 1.7 & \vline & 1.4 & \vline & {\bf 10.26} & \vline & 2.6 & \vline & 1.5\\
140\,kV & \vline & {\bf 7.346} & \vline & 1.3 & \vline & 1.3 & \vline & {\bf 10.15} & \vline & 2.6 & \vline & 1.3\\
 \end{tabular}
 \label{tab:InhFiltrationValues}
 \end{table}

The reproducibility from our previous work \cite{Rosendahl2017} has been preserved, although the procedure has been changed to include a smaller collimation. The uncertainty of the fit has been estimated from an $\chi^2$-analysis of the single measurements, with $1\sigma_{\chi}$-uncertainty being defined as the increase of $\chi^2$ to $\chi^2+1$. From the measurements, the average uncertainty is estimated to be in the range of $1.3\,\%\leq\sigma_{\chi}\leq2.8\,\%$ for the  large and small filters, as it is of the same magnitude as the statistical deviation, as shown in Table \ref{tab:InhFiltrationValues}. 

\begin{figure}[ht] 
   \includegraphics[width=240pt]{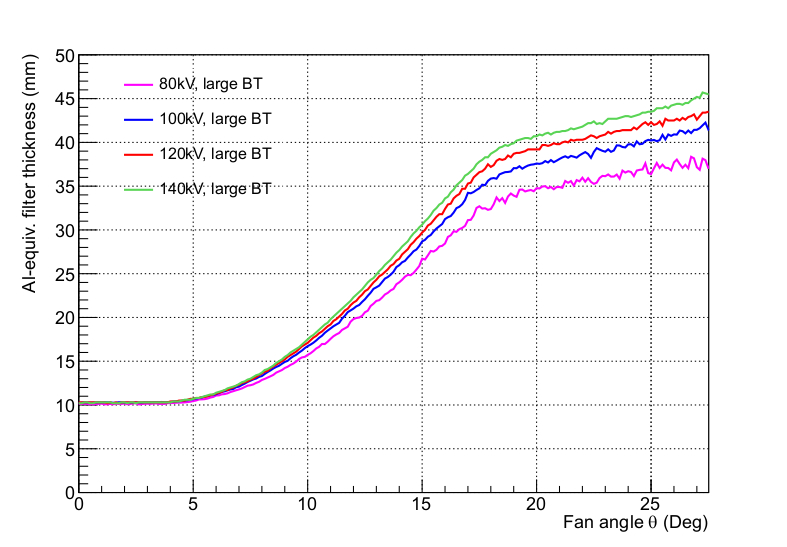} 
	\quad
    \includegraphics[width=240pt]{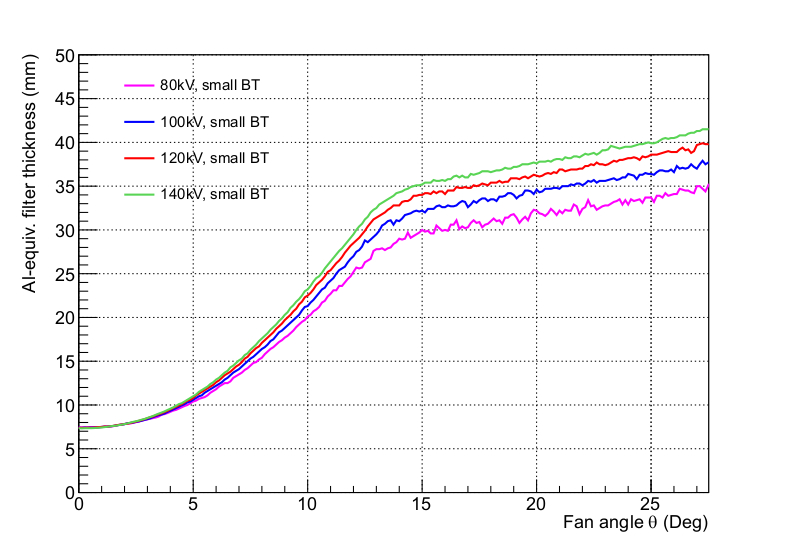} 
		\newline
		\text{ (Large) \hspace{8.0cm} (Small)}
\caption{\textbf{Aluminum equivalent bow tie filters of an Optima 660.} From the eight measurements with the mobile setup, the average aluminum equivalent BT filters for different tube voltages are calculated. The Optima 660 is equipped with a large and a small BT filter.}
\label{fig:EquivalentBT_GE1}
\end{figure}

\begin{figure}[hb] 
   \includegraphics[width=240pt]{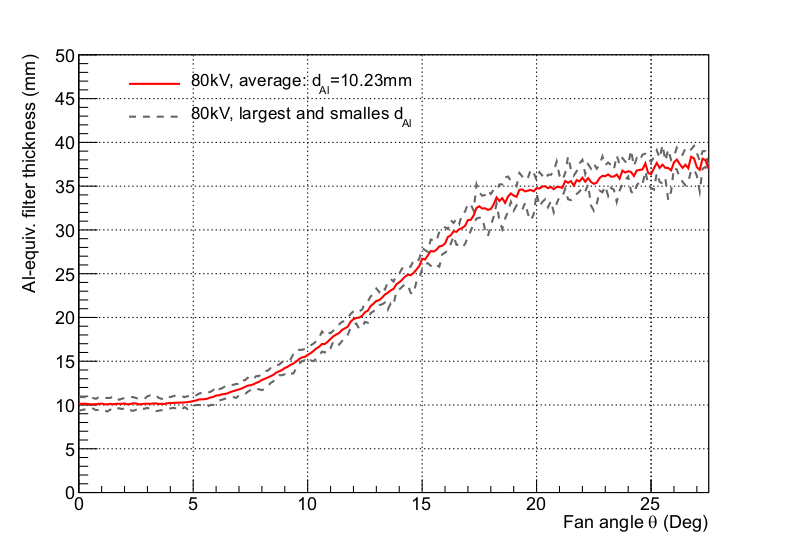} 
	\quad
    \includegraphics[width=240pt]{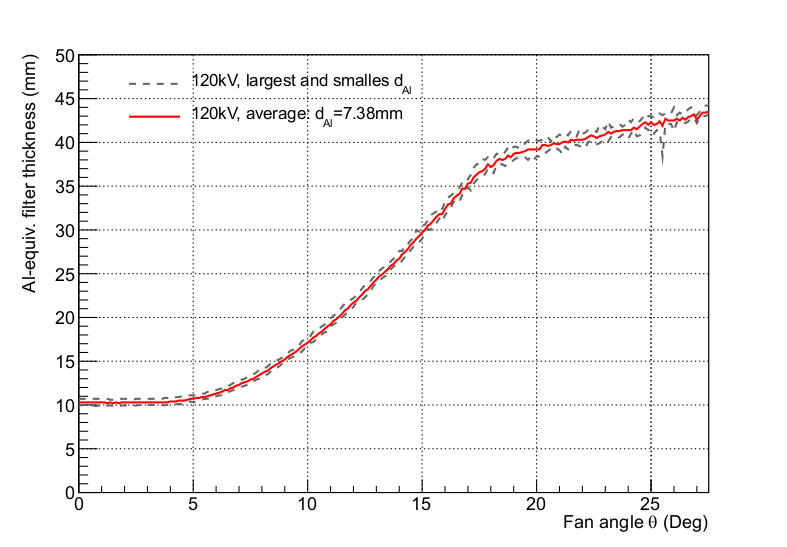} 
		\newline
		\text{ (80kV) \hspace{8.0cm} (120kV)}
\caption{\textbf{Aluminum equivalent large bow tie filter for tube voltages of 80\,kV and 120\,kV.} From the different measurements, the uncertainty in the thickness of the BT filter can be derived and is shown for tube voltages of 80\,kV (left) and 120\,kV (right).}
\label{fig:EquivalentBT_GE2}
\end{figure}

Using the total Al-filtering $d_{{\rm Al}}$, aluminum equivalent BT filter geometries for different tube voltages are constructed for both the small filter and the large filter. From the eight different measurements, the average has been calculated and is shown graphically in Figure \ref{fig:EquivalentBT_GE1}. For large fan angles, a voltage dependent filter thickness appears. This is related to the assumption of only one filter material. In reality, the filter might be constructed from materials other than aluminum, or even from combinations of materials. The energy-dependent photon cross-sections of these materials differ from that of Al, which leads to this spreading. The uncertainty in the determination of the total filtration directly affects the modeling of the equivalent filter; its influence is shown graphically in Figure \ref{fig:EquivalentBT_GE2}. The average large BT filter is plotted for 80\,kV and 120\,kV in comparison to filter geometries with the highest and lowest values of $d_{{\rm Al}}$ obtained from the measurement. The comparison with the standard deviation in Table \ref{tab:InhFiltrationValues} shows that the higher values for 80\,kV are also visible in the spreading of the filter construction, while the deviation for 120\,kV is much smaller.

By way of comparison, additional equivalent source models for the Optima 660 have also been derived using the method of {\bf t}ime-{\bf r}esolved {\bf i}ntegrated {\bf c}harge (TRIC), which is a static approach using the service mode and static X-ray tube. The method and the results of the Optima 660 are presented in \cite{Alikhani2016}. The advantage of the static method is that the measurement conditions can be better controlled and that lead shielding can be applied to improve the collimation and to reduce scattered radiation. Hence, this technique is well suited for benchmarking the dynamic approach, as will be shown in following sections.

\subsubsection{Toshiba Aquilion ONE}

For the determination of the equivalent source model, a collimation of 40\,mm with a pitch of 0.637 is used. The total filtration has been determined under different conditions, and the results are summarized in Table \ref{tab:InhFiltrationValues_Toshiba}. In order to minimize the measurement time in the hospital, not all voltage steps are carried out for each filter. Instead, the Al attenuation measurements are repeated three times for 120\,kV, leading to a standard deviation of $\approx 2\,\%$, which verifies the reproducibility. While the $\chi^2$-analysis of the fitting gives uncertainties of less than 5\,\% in most cases, some single measurements show higher uncertainties of up to 8\,\%. From the values, it is shown that the medium and large filters have the same total thickness at the center of the BT filter, while the small filter is thinner. Within the uncertainty, it is not possible to distinguish between the medium and large filters from the Al-equivalent thickness alone.

\begin{table}[htb]
 \caption{{\bf Total aluminum filtration of the Toshiba Aquilion ONE at the St\"adtisches Klinikum hospital in Braunschweig.} The values for the total Al filtration $d_{{\rm Al}}$ are given for different tube voltages and BT filters.}
\vspace{10pt}
 \begin{tabular}{ccccccccccccc}
  Voltage & \vline &\multicolumn{3}{c}{Small filter}   & \vline &\multicolumn{3}{c}{Medium filter} & \vline &\multicolumn{3}{c}{Large filter}\\
	& \vline & $d_{{\rm Al}}$ [mm]  & \vline & $1\sigma_{\chi}$ [\,\%] & \vline & $d_{{\rm Al}}$ [mm]  & \vline & $1\sigma_{\chi}$ [\,\%] & \vline & $d_{{\rm Al}}$ [mm]  & \vline & $1\sigma_{\chi}$ [\,\%]\\
\hline
\hline
80\,kV & \vline & {\bf 5.880} & \vline & 2.4 & \vline & {\bf -} & \vline & - & \vline & {\bf 7.160} & \vline & 2.7\\
100\,kV & \vline & {\bf 6.120} & \vline & 1.5 & \vline & {\bf -} & \vline & - & \vline & {\bf 7.450} & \vline & 1.3\\
120\,kV & \vline & {\bf 6.143} & \vline & 2.0 & \vline & {\bf 7.901} & \vline & 4.9 & \vline & {\bf 7.540} & \vline & 1.8\\
135\,kV & \vline & {\bf 6.220} & \vline & 7.6 & \vline & {\bf -} & \vline & - & \vline & {\bf 7.875} & \vline & 8.1\\

 \end{tabular}
 \label{tab:InhFiltrationValues_Toshiba}
 \end{table}

\begin{figure}[ht]
   \includegraphics[width=240pt]{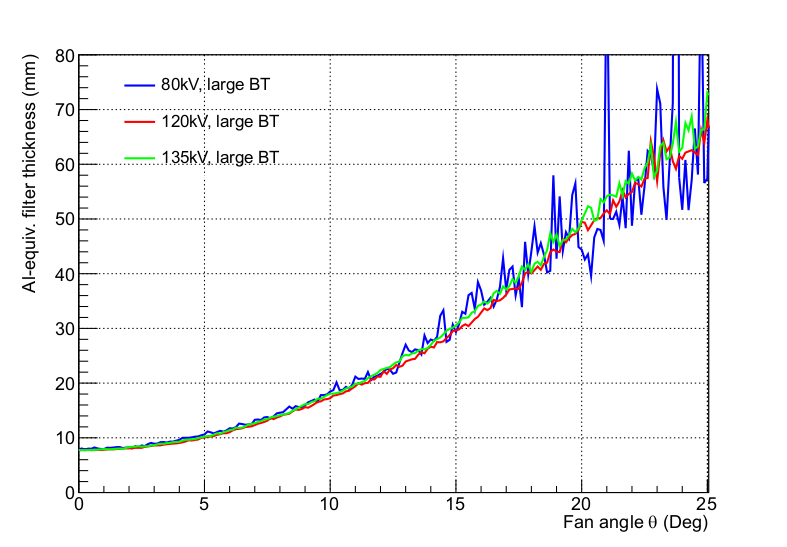} 
	\quad
    \includegraphics[width=240pt]{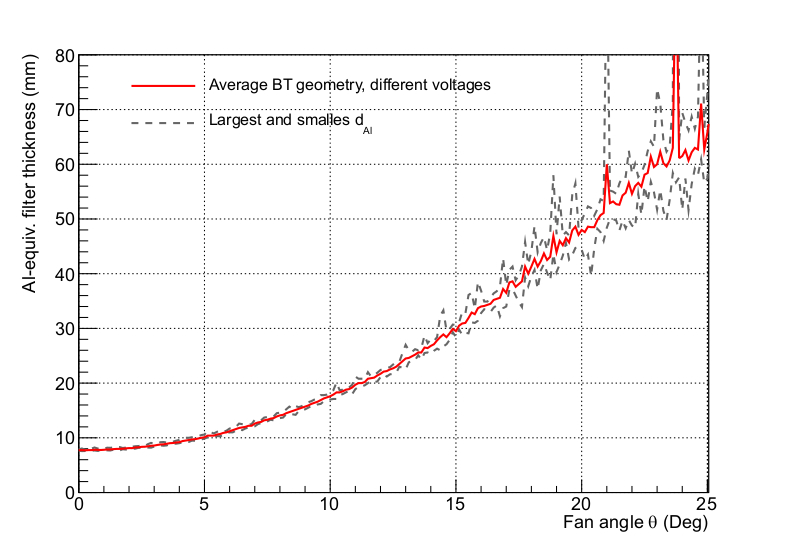} 
		\newline
		\text{ (a) \hspace{8.0cm} (b)}
\caption{\textbf{Aluminum equivalent bow tie filters of a Toshiba Aquilion ONE.} The mobile setup has been used to evaluate the Al-equivalent BT filter geometry of the large filter and for different tube voltages (a). From the single measurements, an average filter for dose simulations is calculated (b).}
\label{fig:AquillionONE_EqBT_Al}
\end{figure}

Since the large BT filter is used for the validation process in this work, we focus on the determination of the equivalent thickness of this filter while the results for the other filters are planned for publication elsewhere. In Figure \ref{fig:AquillionONE_EqBT_Al}, the Al-equivalent geometry of the large filter is presented for different tube voltages (left). In contrast to the results from the Optima 660 shown in Figure \ref{fig:EquivalentBT_GE1}, the geometry does not split for the different tube voltages at large fan angles. Hence, it can be assumed that the filter is made from materials whose attenuation characteristics are similar to aluminum. Consequently, an average filter from the different tube voltages can be calculated. 

Compared to the Optima 660, the filtration at the center of the BT is lower, while at the edges, the opposite is true. This should lead to higher values for air kerma on-axis at the iso center. For the 80\,kV filter geometry, some conspicuous peaks occur at high fan angles. These artifacts are related to the low signal intensity of $\mathcal{O}(5\cdot 10^{-3}\,{\rm mGy/s})$ due to the high filtering at large fan angles. This effect has not been observed at the Optima 660 since the absolute thickness of the filter in the Aquilion ONE is higher at the edges compared to the Optima 660.

\subsection{Validation of equivalent source models with dose measurements and simulations.}
\label{sec:AQUILION_One}

\subsubsection{GE Optima CT 660}

For the validation, different simulations have been performed using equivalent source models obtained from the dynamic method and the TRIC method. In this work, the results for tube voltages of 80\,kV and 120\,kV at a nominal collimation of 10\,mm are presented for the Body PMMA CT phantom and the thorax phantom in order to provide a manageable number of values. For the proper normalization, the air kerma at the iso center has been measured free-in-air as described above. The values for the $CTDI_{{\rm free\, air}}$ and the peak air kerma, $K\rm{_{a}^{P}}$, as well as the total beam width $ z_{{\rm Tot}}$ for the BT filter and tube voltages used in the Optima 660 are presented in Table \ref{tab:K0}. The estimated uncertainty budget for air kerma measurements are compiled in Table \ref{tab:UncertaintyBudget_GE} for the different detectors. The relative uncertainty of $z_{{\rm Tot}}$ is calculated as 1.2\,\% for $k=1$; its influence on the simulation results is investigated below. 

\begin{table}[hb]
 \caption{{\bf CT dose output of the Optima 660.} The values for $CTDI_{{\rm free\, air}}$ and the average air kerma $K\rm{_{a}^{P}}$ are measured using the 100\,mm pencil-type ionization chamber and the RC0.6, respectively. The values are used to calculate the total beam width $z_{{\rm Tot}}$. The combined uncertainty $\sigma_{\rm c}$ in the dose measurement and total beam width was calculated for a coverage factor of $ k=1$.}
\vspace{8pt}
 \begin{tabular}{ccccccccccccccc}
  & \vline &  & \vline &   &\vline &\multicolumn{3}{c}{10\,mm}   & \vline &\multicolumn{3}{c}{40\,mm} \\
	
Filter & \vline &	Voltage & \vline  & $K\rm{_{a}^{P}}$ &\vline &$CTDI_{{\rm free\, air}}$ &  \vline & $z_{{\rm Tot}}$ & \vline & $CTDI_{{\rm free\, air}}$ &  \vline & $z_{{\rm Tot}}$ \\
& \vline &		& \vline & [mGy/100\,mAs] &\vline & [mGy/100\,mAs] &  \vline & [mm] & \vline & [mGy/100\,mAs]  & \vline & [mm] \\
& \vline &		& \vline & $\sigma_{\rm c}=0.7\,\% $ &\vline & $\sigma_{\rm c}=1.0\,\%$ &  \vline & $\sigma_{\rm c}=1.2\,\%$ & \vline & $\sigma_{\rm c}=1.0\,\%$ & \vline & $\sigma_{\rm c}=1.2\,\%$ \\

\hline
\hline
Large & \vline & 80\,kV & \vline & 7.575 &\vline &  10.10 & \vline &  13.3   & \vline &   8.496    & \vline &  44.9 \\
& \vline & 120\,kV & \vline &  21.30 & \vline &  28.26  & \vline & 13.3 & \vline &   23.74   & \vline &  44.5  \\

 \end{tabular}
 \label{tab:K0}
 \end{table}

Air kerma measurements are performed inside the homogenous PMMA body phantom using the axial scan mode with only one rotation of 1\,s and a current of 100\,mA. Five calibrated dose profilers were used for the measurement. In Table \ref{tab:PMMA_Point}, the results from the measurement and from the simulation are compiled.

\begin{table}[ht]
 \caption{{\bf Results of point dose measurements and simulations performed on the Optima 660.} The dose values at the five positions of the PMMA 32\,cm phantom are measured with semiconductor detectors (RTI) and compared to simulated values from ImpactMC. The scan parameters chosen are one rotation with a current of 100\,mA a rotation time of 1\,s and a nominal collimation of 10\,mm.}
\vspace{10pt}
 \begin{tabular}{ccccccccccccccc}
 
	 &\vline &  & \vline &  & \vline& Measured dose & \vline &  \multicolumn{5}{c}{Simulated dose} \\
	Voltage	 & \vline & Position & \vline  & $k_{{\rm Q,Ph}}$ & \vline  & [mGy/100\,mAs] & \vline & Dynamic method & \vline &  Difference & \vline &  TRIC method & \vline &  Difference\\
 & \vline &  & \vline  & & \vline  & $\sigma_{\rm c}=1.3\,\% $ $(k=1)$ & \vline & [mGy/100\,mAs]  & \vline &  [\,\%] & \vline &  [mGy/100\,mAs] & \vline & [\,\%]\\		
 
 \hline

 & \vline & 12 o'c & \vline & 0.9699 & \vline  & {\bf 2.789} & \vline  & {\bf 2.632} & \vline & -5.6 & \vline & {\bf 2.783} & \vline & -0.2\\
 & \vline & 3 o'c & \vline & 0.9657& \vline  & {\bf 2.611} & \vline  & {\bf 2.588} & \vline & -0.9 & \vline & {\bf 2.754} & \vline & 5.5\\
80\,kV & \vline & 6 o'c & \vline & 0.9671& \vline  & {\bf 2.035} & \vline  & {\bf 1.953} & \vline & -4.1 & \vline & {\bf 2.172} & \vline & 6.7\\
 & \vline & 9 o'c & \vline & 0.9785& \vline  & {\bf 2.629} & \vline  & {\bf 2.708} & \vline & 3.0 & \vline & {\bf 2.804} & \vline & 6.7\\
 & \vline & center & \vline & 0.9830& \vline  & {\bf 0.4129} & \vline  & {\bf 0.303} & \vline & 4.2 & \vline & {\bf 0.4265} & \vline & 3.3\\

\hline

 & \vline & 12 o'c & \vline & 0.9384& \vline  & {\bf 8.178} & \vline  & {\bf 7.997} & \vline & -2.2& \vline & {\bf 8.326} & \vline & 1.8\\
 & \vline & 3 o'c & \vline & 0.9352& \vline  & {\bf 7.682} & \vline  & {\bf 7.895} & \vline & 2.8 & \vline & {\bf 8.094} & \vline & 5.4\\
120\,kV & \vline & 6 o'c & \vline & 0.9369& \vline  &  {\bf 6.193} & \vline  & {\bf 6.246} & \vline & 0.9 & \vline & {\bf 6.437} & \vline & 3.9\\
 & \vline & 9 o'c & \vline & 0.9445& \vline  & {\bf 7.855} & \vline  & {\bf 8.025} & \vline & 2.2 & \vline & {\bf 8.318} & \vline & 5.9\\
 & \vline & center & \vline & 0.9131& \vline  & {\bf 1.511} & \vline  & {\bf 1.632} & \vline & 8.0 & \vline & {\bf 1.593} & \vline & 5.4\\

 \end{tabular}
 \label{tab:PMMA_Point}
 \end{table}

The mean average difference between the measured and the simulated doses is determined to be $3.6\,\%$ for 80\,kV and $3.2\,\%$ for 120\,kV with a maximum difference of 5.6\,\% and 8.0\,\%, respectively; the dynamic determination of equivalent source models show good agreement. The combined uncertainty for the measurement of the air kerma with the dose profilers has been estimated as $\sigma_{\rm c}=1.3\,\% $ for a coverage factor $k=1$, using the maximum deviation observed from all positions and voltages after three repeated measurements of $\sigma_{{\rm rep, max}}=0.75\,\%$.  Hence, the dose measurement and the simulation agree at all positions within the expanded uncertainty for $k=2$.

\begin{table}[ht]
 \caption{{\bf Results of dose measurements with 100\,mm pencil-type chamber and simulations performed on the Optima 660.} The air kerma length product $K_{\rm a} L$ at the five positions of the PMMA 32\,cm phantom are measured with a 100\,mm CT pencil-type chamber (PTW) and compared to simulated values from ImpactMC. The scan parameters chosen are one rotation with a current of 100\,mA, a rotation time of 1\,s and a nominal collimation of 10\,mm.}
\vspace{10pt}
 \begin{tabular}{ccccccccccccc}
 
	Voltage &\vline & Position & \vline & Measured $K_{\rm a} L$ & \vline &  \multicolumn{5}{c}{Simulated $K_{\rm a} L$} \\
		 & \vline &  & \vline  & [mGy$\cdot$cm/100\,mAs]  & \vline & Dynamic method & \vline &  Difference & \vline &  TRIC method & \vline &  Difference\\
 & \vline &  & \vline  & $\sigma_{\rm c}=1.3\,\%$ & \vline & [mGy$\cdot$cm/100\,mAs]  & \vline &  [\,\%] & \vline &  [mGy$\cdot$cm/100\,mAs] & \vline & [\,\%]\\		
 \hline

 & \vline & 12 o'c & \vline & {\bf 4.457} & \vline  & {\bf 4.229} & \vline & -5.1 & \vline & {\bf 4.265} & \vline & -4.3\\
 & \vline & 3 o'c & \vline & {\bf 4.359} & \vline  & {\bf 4.163} & \vline & -4.5 & \vline & {\bf 4.217} & \vline & 0-3.3\\
80\,kV & \vline & 6 o'c & \vline & {\bf 3.835} & \vline  & {\bf 3.625} & \vline & -5.5 & \vline & {\bf 3.774} & \vline & -1.6\\
 & \vline & 9 o'c & \vline & {\bf 4.373} & \vline  & {\bf 4.234} & \vline & -3.2 & \vline & {\bf 4.231} & \vline & -3.2\\
 & \vline & center & \vline & {\bf 1.709} & \vline  & {\bf 1.884} & \vline & 10.2 & \vline & {\bf 1.871} & \vline & 9.5\\

\hline

 & \vline & 12 o'c & \vline & {\bf 13.43} & \vline  & {\bf 12.75} & \vline & -5.1 & \vline & {\bf 12.59} & \vline & -6.3\\
 & \vline & 3 o'c & \vline & {\bf 13.23} & \vline  & {\bf 12.674} & \vline & -3.7 & \vline & {\bf 12.43} & \vline & -6.1\\
120\,kV & \vline & 6 o'c & \vline & {\bf 11.89} & \vline  & {\bf 10.99} & \vline & -7.5 & \vline & {\bf 11.34} & \vline & -4.6\\
 & \vline & 9 o'c & \vline & {\bf 13.26} & \vline  & {\bf 12.64} & \vline & -4.6 & \vline & {\bf 12.45} & \vline & -6.1\\
 & \vline & center & \vline & {\bf 6.425} & \vline  & {\bf 6.876} & \vline & 7.0 & \vline & {\bf 6.565} & \vline & 2.2\\

 \end{tabular}
 \label{tab:PMMA_DLP100}
 \end{table}

Furthermore, the air kerma length product is measured using a 100\,mm pencil-type chamber under the same conditions, with the results compiled in Table \ref{tab:PMMA_DLP100}. For the measurements of the air kerma length product, the combined uncertainty is estimated to be $\sigma_{\rm c}=1.1\,\% $ for coverage factor $(k=1)$, using the maximum deviation observed from all positions and voltages after three repetitions of the measurements of $\sigma_{{\rm rep, max}}=0.7\,\%$. The average differences between the simulation and the measurement are found to be 5.7\,\% for 80\,kV and 5.6\,\% 120\,kV with a maximum difference of 10.2\,\% and 7\,\%, respectively. This is in agreement with the measurements of the point dose with the semiconductor detector. For the 6\,\,o'clock position, the influence of the patient table is significant, without an equivalent table material being defined. While the scan field of view is optimized to achieve the best resolution for the anatomy of the patient, typically, only parts of the table structure are covered and included in the image file. Hence, the table structure is not available for simulations and needs to be added artificially to achieve accurate simulation results.

The TRIC model yields similar results for the measurement of the air kerma length product. In the margin of uncertainty, the values are comparable, indicating that the dynamic approach is viable and can be used to determine equivalent source models.

\begin{table}[ht]
 \caption{{\bf Results of point dose measurements and simulations in an anthropomorphic phantom performed on the Optima 660.} The dose values at the five positions of the ``Thorax-Medium Adult'' anthropomorphic phantom are measured with semiconductor detectors (RTI) and compared to simulated values from ImpactMC. The scan parameters chosen are one rotation with 100\,mA, a rotation time of 1\,s and a nominal collimation of 10\,mm.}
\vspace{10pt}
 \begin{tabular}{ccccccccccccc}

	 &\vline &  & \vline & Measured & \vline &  \multicolumn{3}{c}{Simulated dose} \\
	Voltage	 & \vline & Position & \vline  &dose & \vline & Dynamic method & \vline &  Difference \\
 & \vline &  & \vline  &  [mGy/100\,mAs]& \vline & [mGy/100\,mAs]  & \vline &  [\,\%] \\		
 
 \hline
\hline

 & \vline & 12 o'c & \vline & {\bf 2.889} & \vline  & {\bf 2.813} & \vline & -2.6 \\
 & \vline & 3 o'c & \vline & {\bf 2.437} & \vline  & {\bf 2.426} & \vline & -0.4 \\
80\,kV & \vline & 6 o'c & \vline & {\bf 2.204} & \vline  & {\bf 2.254} & \vline & 2.3 \\
 & \vline & 9 o'c & \vline & {\bf 2.426} & \vline  & {\bf 2.335} & \vline & -3.7 \\
 & \vline & center & \vline & {\bf 0.7085} & \vline  & {\bf 0.7067} & \vline & -0.2 \\

\hline

 & \vline & 12 o'c & \vline & {\bf 8.451} & \vline  & {\bf 8.791} & \vline & 4.0 \\
 & \vline & 3 o'c & \vline & {\bf 7.169} & \vline  & {\bf 7.394} & \vline & 3.1 \\
120\,kV & \vline & 6 o'c & \vline & {\bf 6.627} & \vline  & {\bf 6.963} & \vline & 5.1 \\
 & \vline & 9 o'c & \vline & {\bf 7.407} & \vline  & {\bf 7.291} & \vline & -1.6 \\
 & \vline & center & \vline & {\bf 2.545} & \vline  & {\bf 2.692} & \vline & 5.8 \\
 
 \end{tabular}
 \label{tab:THXAdult_Point}
 \end{table}

In the next step, point dose measurements with semiconductor sensors are performed in an anthropomorphic phantom using the ``Thorax-Medium Adult'' phantom. The  scan parameters are set to 100\,mA, one rotation and a rotation time of 1\,s, which are the same as in previous measurements. The material conversion information for lungs, bones and water from the ImpactMC manual \cite{ABCT2014} were used, which yield the simulated dose values compiled in Table \ref{tab:THXAdult_Point}. In this example, the simulated doses are larger than those measured using the RTI dose profilers. For the 80\,kV data, the average difference observed is 1.9\,\%, with a maximum difference of 3.7\,\%. For the 120\,kV measurement, the average difference is slightly higher, being 3.9\,\% with a maximum difference of 5.8\,\% in the central position. For the air kerma measurements, a combined uncertainty has been estimated to be $\sigma_{\rm c}=2.0\,\% $ for a coverage factor $(k=1)$, using the maximum deviation observed from all positions and voltages after three repeated measurements of $\sigma_{{\rm rep, max}}=1.7\,\%$.

In order to validate the concept for scan parameters that are closer to those used in clinical practice, the thorax phantom is scanned with the ``routine chest 0.6 s 5mm SMARTmA'' GE protocol. This helical scan has a pitch of 1.375 and a source rotation time of 0.6\,s. The voltage is set to 120\,kV and the large BT filter is used. Although this scan protocol normally uses a collimation of 40\,mm, a smaller collimation of 20\,mm is chosen due to the shortness of the phantom. The dose measurement is improved and additionally the uncertainties from the edges of the phantom are reduced. Furthermore, the ``SMARTmA'' current modulation technique, which adjusts the current to the patient, was switched off to simplify the simulation. The current value has been fixed to 300\,mA for the whole scan range of 6\,cm, resulting in a measurement time of 2.01\,s.

\begin{figure}[hb]
   \includegraphics[width=150pt]{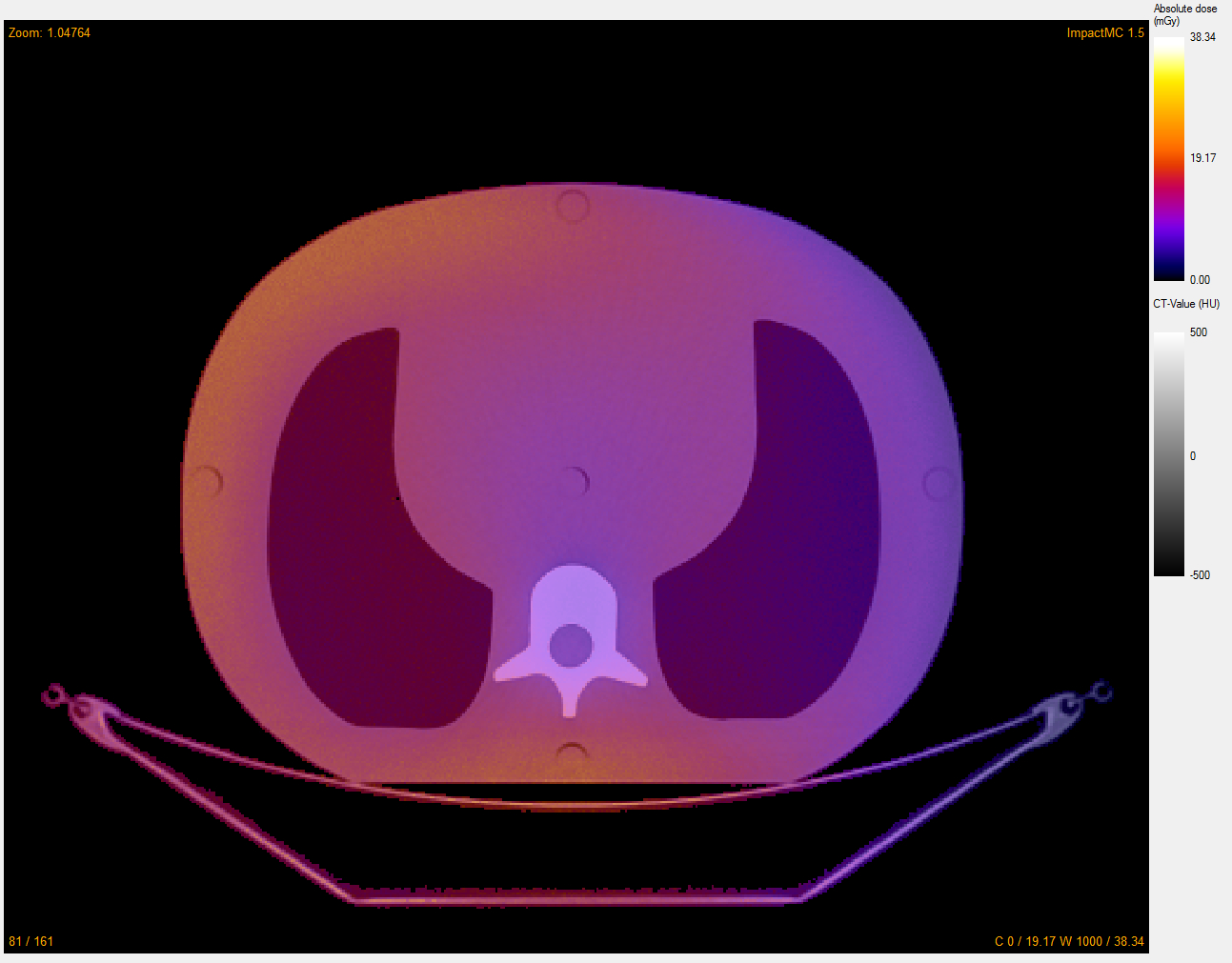} 
	\quad
    \includegraphics[width=150pt]{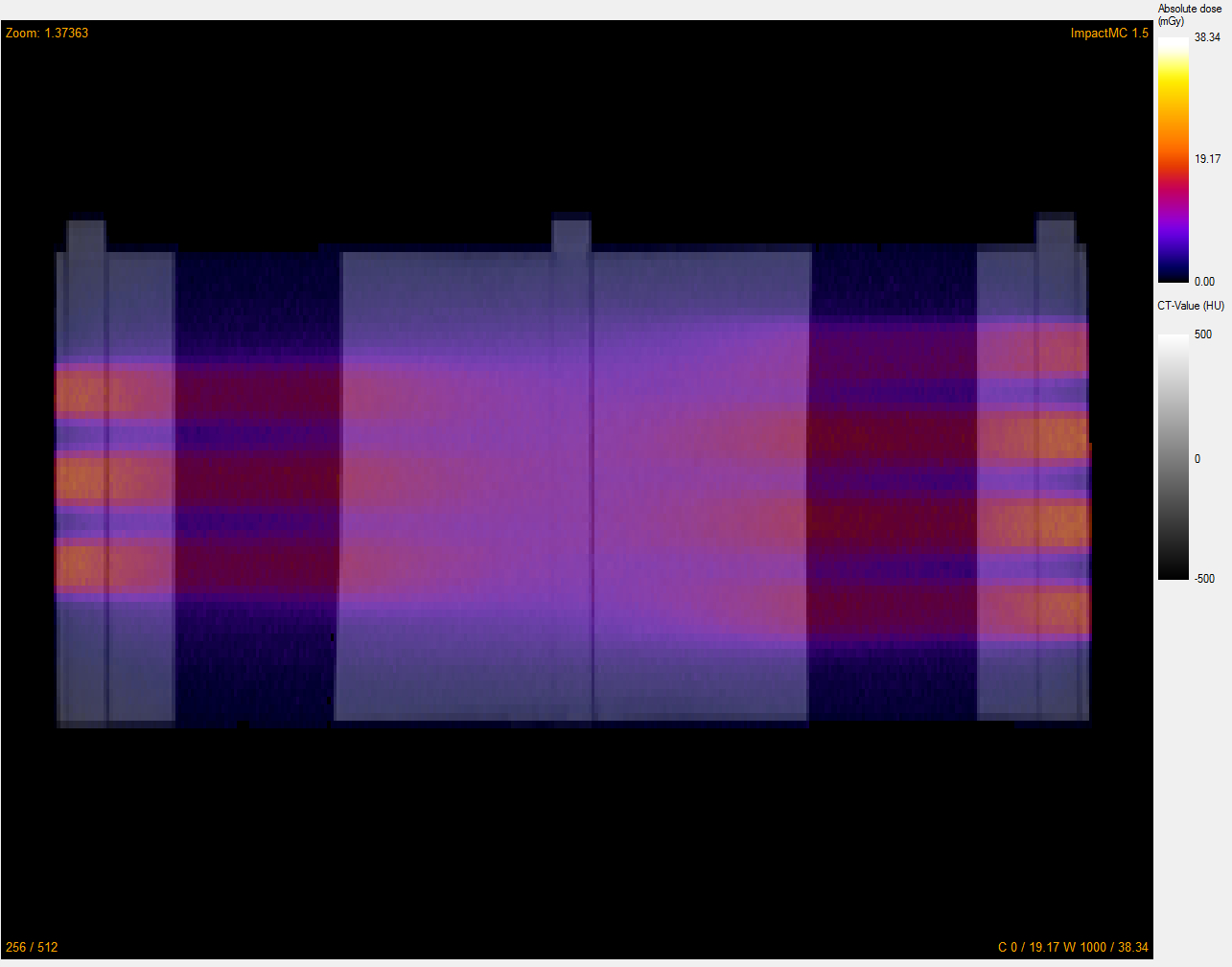} 
	\quad
    \includegraphics[width=150pt]{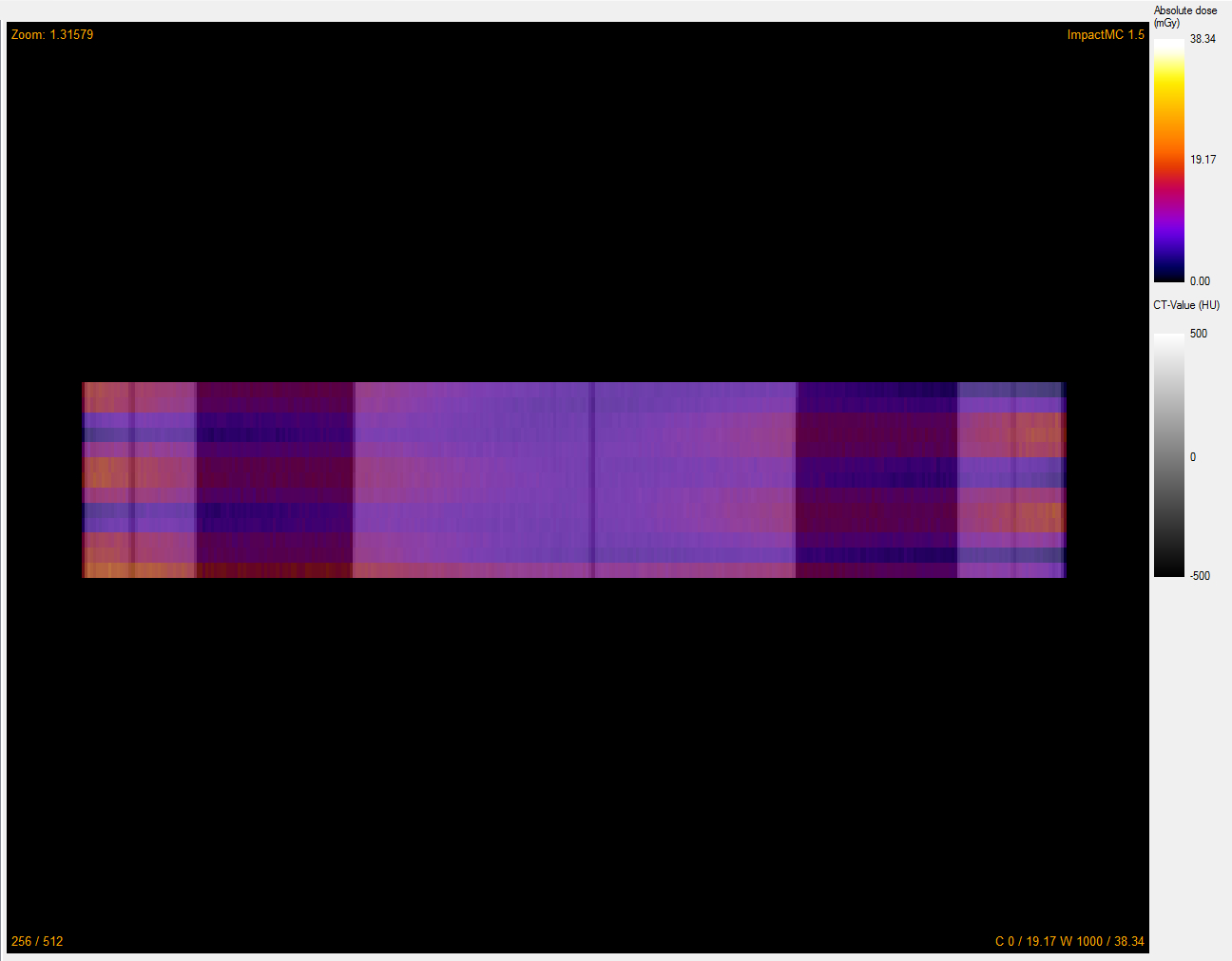}
		\newline
		\text{ (a) \hspace{4.5cm} (b) \hspace{5.2cm} (c)}
\caption{\textbf{Dose simulation of a routine chest protocol (GE).} A standard patient protocol provided by General Electric, ``routine chest 0.6 s 5mm SMARTmA'', is applied to the ``Thorax-Medium Adult'' anthropomorphic phantom and the dose simulation is performed using the equivalent source model of the Optima 660. The 3D dose distribution is shown for full geometry of the phantom (a+b) and for the 6\,cm scan length (c), which has been defined as the scan area. The images in (b+c) show the projection in z-direction (``Coronal'' window of ImpactMC).}
\label{fig:Sim_ChestStd}
\end{figure}

The air kerma length product for the exposure is measured with a pencil-type ionization chamber and an Unidos electrometer readout. In Figure \ref{fig:Sim_ChestStd}, the ImpactMC user display for the 3D dose simulation is shown. 

Although only 6\,cm of the phantom are reserved for projection, a total scan length of $\approx 9.2\,$cm is calculated from the $CTDI_{{\rm vol}}$ and $DLP$ values given by the system (see Table \ref{tab:THXAdult_DLP_ChestStd}). The difference is caused by the well-known over-beaming effect related to the spiral scan modality. Hence, to achieve good dose results, the complete geometry of the phantom is needed for the simulation. Otherwise, the simulated dose will be drastically underestimated in the way it is illustrated in Figure \ref{fig:Sim_ChestStd}, where the dose maps (b) and (c) are compared. For the studies in this work, the whole phantom and additional space before and after were scanned to address possible scattering from the table. In reality, because only a part of the patient is scanned, the border areas need to be modeled. The results from the simulation and from the measurement of the air kerma length product are compiled in Table \ref{tab:THXAdult_DLP_ChestStd}. 

\begin{table}[ht]
 \caption{{\bf Results of measurements of the air kerma length product and simulations in an anthropomorphic phantom using routine chest protocol.} The air kerma length product $K_{\rm a} L$ at the five positions of the ``Thorax-Medium Adult'' anthropomorphic phantom are measured with a pencil-type ionization chamber (100\,mm) and are compared to simulated values from ImpactMC. A routine chest protocol with a 5\,mm slice thickness and 120\,kV in the Optima 660 is used.}
\vspace{10pt}
 \begin{tabular}{ccccccccccccc}
 
	 &\vline &  & \vline & Measured $K_{\rm a} L$& \vline &  Simulated $K_{\rm a} L$ & \vline &\\
	Voltage	 & \vline & Position & \vline  &[mGy*cm] & \vline & Dynamic method & \vline & Differences \\
 & \vline &  & \vline  &  $\sigma_{\rm c}=3.3\,\% $ & \vline & [mGy*cm]  & \vline & [\,\%]\\		
 
 \hline

 & \vline & 12 o'c & \vline & { 147.2} & \vline  & { 150.4} & \vline & 2.2 \\
 & \vline & 3 o'c & \vline & { 116.2} & \vline  & { 118.4} & \vline & 1.9 \\
120\,kV & \vline & 6 o'c & \vline & { 125.1} & \vline  & { 129.5} & \vline & 3.5 \\
 & \vline & 9 o'c & \vline & { 119.0} & \vline  & { 120.5} & \vline & 4.3 \\
 & \vline & center & \vline & { 84.72} & \vline  & { 88.97} & \vline & 5.0 \\

 \end{tabular}
 \label{tab:THXAdult_DLP_ChestStd}
 \end{table}

The combined uncertainty of the measurement has been determined to be $\sigma_{\rm c}=2.7\,\%$ and is dominated by the statistical uncertainty that occurs during the measurement of $\sigma_{{\rm Rep}}=2.5\,\%$. The measured and simulated values agree within the uncertainties of the measurements and the simulation. The average difference is calculated to be 3.4\,\% with a maximum difference of 5\,\%. For this simulation, the starting angle, which is unknown, has been varied between $0^{\circ}$ and $270^{\circ}$ and the average of these simulations is compiled in Table \ref{tab:THXAdult_DLP_ChestStd}. This reduces the uncertainty in the starting angle for helical scans from 7.5\,\% down to 3.9\,\% for a coverage factor $k=1$. In summary, the test validates the notion that more realistic scan settings can also be simulated with small differences between the measurement and the simulation. 

\subsubsection{Toshiba Aquilion ONE}

For the Aquilion ONE, the air kerma $K\rm{_{a}^{P}}$ and the $CTDI_{{\rm free\, air}}$ were measured with the RC0.6 chamber and with a 300\,mm pencil-type chamber, respectively. In Table \ref{tab:K0_Toshiba}, the results from the measurements and the corresponding values for the total beam collimation are compiled. The absolute values for $K\rm{_{a}^{P}}$ and for $CTDI_{{\rm free\, air}}$ are higher than for the GE Optima CT under comparable conditions, which is expected due to the lower total filtration from the equivalent source model.

\begin{table}[ht]
 \caption{{\bf CT dose output of the Toshiba Aquilion ONE.} The values for the $CTDI_{{\rm free\, air}}$ and the average air kerma $\rm{K_{a}^{P}}$ are measured using a 300\,mm pencil-type chamber and the RC0.6, respectively. The values are used to calculate the total beam width $z_{{\rm Tot}}$. The combined uncertainty $\sigma_{c}$ in the dose measurement and total beam width is calculated using the GUM workbench for a coverage factor $k=1$.}
\vspace{8pt}
 \begin{tabular}{ccccccccccccccccc}
    & \vline &   &\vline &\multicolumn{3}{c}{12\,mm}   & \vline &\multicolumn{3}{c}{40\,mm} & \vline &\multicolumn{3}{c}{160\,mm}\\
	
	Voltage & \vline  & $K{\rm_{a}^{P}}$ &\vline &$CTDI_{{\rm free\, air}}$ &  \vline & $z_{{\rm Tot}}$ & \vline & $CTDI_{{\rm free\, air}}$ &  \vline & $z_{{\rm Tot}}$ & \vline & $CTDI_{{\rm free\, air}}$ &  \vline & $z_{{\rm Tot}}$\\
		& \vline & [mGy/100\,mAs] &\vline & [mGy/100\,mAs] &  \vline & [mm] & \vline & [mGy/100\,mAs]  & \vline & [mm] & \vline & [mGy/100\,mAs]  & \vline & [mm]\\
		& \vline & $\sigma_{\rm c}=0.7\,\% $ &\vline & $\sigma_{\rm c}=1.1\,\%$ &  \vline & $\sigma_{\rm c}=1.3\,\%$ & \vline & $\sigma_{\rm c}=1.1\,\%$ & \vline & $\sigma_{\rm c}=1.3\,\%$ & \vline & $\sigma_{\rm c}=1.1\,\%$ & \vline & $\sigma_{\rm c}=1.3\,\%$\\
\hline
\hline
 80\,kV & \vline & 9.670 &\vline &  12.85 & \vline &  15.94   & \vline &   10.83    & \vline &  44.29  & \vline &   9.866   & \vline &  163.3\\
 120\,kV & \vline & 25.02 &\vline &  32.60 & \vline &  15.63   & \vline &   27.71    & \vline &  44.78  & \vline &   25.48    & \vline &  163.0\\

 \end{tabular}
 \label{tab:K0_Toshiba}
 \end{table}

For the validation of the equivalent source model, the measured air kerma length product is compared to a simulation in an anthropomorphic phantom. In Table \ref{tab:THXAdult_DLP_AqONE}, the results for one tube rotation at 120\,kV tube voltage are shown. It has been observed that, after repeating the measurement three times, the values of measured air kerma length product using the pencil-type ionization chamber can vary up to 8.5\,\%, leading to a very high combined uncertainty of $\sigma_{\rm c}\approx 8.6\,\%$ ($k=1$) in the dose measurement. The reason for the fluctuation is related to the rotation angle of the tube when starting the irradiation, which varies over time.  The effect is much more pronounced for the Toshiba Aquilion ONE than for the GE Optima 660. This fact may be due to the construction of the CT.

\begin{table}[ht]
 \caption{{\bf Results of measurements and simulations of the air kerma length product in anthropomorphic phantom at the Toshiba Aquilion ONE.} The air kerma at the five positions of the ``Thorax-Medium Adult'' anthropomorphic phantom are measured with a 100\,mm pencil-type ionization chamber (PTW) and compared to simulated values from ImpactMC. The scan parameters have been set to one rotation with 150\,mA a rotation time of 1\,s and a nominal collimation of 12\,mm.}
\vspace{10pt}
 \begin{tabular}{ccccccccc}

	 &\vline &  & \vline & Measured $K_{\rm a} L$& \vline &  Simulated $K_{\rm a} L$ & \vline &\\
	Voltage	 & \vline & Position & \vline  &[mGy$\cdot$cm/100\,mAs] & \vline & Dynamic method & \vline & Differences \\
 & \vline &  & \vline  &  $\sigma_{\rm c}=3.3\,\% $ & \vline & [mGy$\cdot$cm/100\,mAs]  & \vline & [\,\%]\\		
 
 \hline

 & \vline & 12 o'c & \vline & { 18.23} & \vline  & { 18.21} & \vline & -0.2 \\
 & \vline & 3 o'c & \vline & { 14.29} & \vline  & { 14.54} & \vline & 1.7 \\
120\,kV & \vline & 6 o'c & \vline & {14.71} & \vline  & { 15.60} & \vline & 6.1 \\
 & \vline & 9 o'c & \vline & { 14.42} & \vline  & {14.65} & \vline & 1.6 \\
 & \vline & center & \vline & { 10.66} & \vline  & {11.63} & \vline & 9.2 \\

 \end{tabular}
 \label{tab:THXAdult_DLP_AqONE}
 \end{table}

The average difference between the measurement and the simulation has been determined to be 3.8\,\% with the maximum difference of 9.2\,\%. The overall results are in agreement with the studies performed on the Optima 660, indicating a slight overestimation of the air kerma in the simulations for the 6 o'clock and  central positions of the phantom.

Finally, more realistic scan settings are presented. The anthropomorphic phantom is scanned using a spiral scan mode with a pitch of 0.813 and a nominal collimation of 40\,mm, covering a scan range of 8\,cm around the center of the phantom. The tube voltage was set to 120\,kV with a constant current of 100\,mA without the tube current modulation. The air kerma length product measured using the 100\,mm pencil-type ionization chamber together with the simulation results are compiled in Table \ref{tab:THXAdult_AqONE_Spiral}.

\begin{table}[ht]
 \caption{{\bf Results of measurements and simulations of the air kerma length product in an anthropomorphic phantom at the Toshiba Aquilion ONE using a spiral scan mode.} The dose values at the five positions of the ``Thorax-Medium Adult'' anthropomorphic phantom are measured with a 100\,mm pencil-type ionization chamber (PTW) and compared to simulated values from ImpactMC. The scan parameters have been set to spiral mode, covering 8\,cm with 100\,mA a rotation time of 0.5\,s and a nominal collimation of 40\,mm.}
\vspace{10pt}
 \begin{tabular}{ccccccccc}

	 &\vline &  & \vline & Measured $K_{\rm a} L$& \vline &  Simulated $K_{\rm a} L$ & \vline &\\
	Voltage	 & \vline & Position & \vline  &[mGy$\cdot$cm/100\,mAs] & \vline & Dynamic method & \vline & Differences \\
 & \vline &  & \vline  &  $\sigma_{\rm c}=3.3\,\% $ & \vline & [mGy$\cdot$cm/100\,mAs]  & \vline & [\,\%]\\		
 
 \hline

 & \vline & 12 o'c & \vline & { 75.20} & \vline  & { 76.17} & \vline & 7.4 \\
 & \vline & 3 o'c & \vline & { 61.49} & \vline  & { 63.03} & \vline & 1.3 \\
120\,kV & \vline & 6 o'c & \vline & {63.85} & \vline  & { 66.54} & \vline & 4.2 \\
 & \vline & 9 o'c & \vline & { 61.02} & \vline  & {66.54} & \vline & -1.5 \\
 & \vline & center & \vline & { 46.00} & \vline  & {49.42} & \vline & 7.4 \\

 \end{tabular}
 \label{tab:THXAdult_AqONE_Spiral}
 \end{table}

The results of the measurement and the simulation show differences of less than 8\,\% with an average difference of 4.4\,\% over all positions. During this measurement, the uncertainty from repetition (three times) was very small, on the order of $\sigma_{{\rm Rep}}\approx 1.0\,\%$ ($k=1$). Hence, the combined uncertainty for the dose measurement is calculated to be $\sigma_{{\rm c, meas}}=1.4\,\% $ for coverage factor $(k=1)$ using the systematic uncertainties (Type B) from Table \ref{tab:UncertaintyBudget_GE}. The reason for the large difference in the statistical uncertainty in the tube output between measurements with one rotation and measurements with several rotations is unknown.

\section{Discussion of the results}
\label{sec:Discussion}

In summary, the measured and simulated values for the Optima 660 in both standard CTDI and anthropomorphic phantoms using equivalent source models from a dynamic determination agree within a few percent, with only a few examples of larger differences. In different measurements, it was observed that the simulated dose in the central position was overestimated in the simulation, which indicates that the value for total filtration $d_{{\rm Al}}$ is slightly overestimated, thus causing a harder spectrum. The latest measurements using static methods with lead shielding of scattered radiation from the X-ray tube would seem to confirm this theory. Alternatively, the phantom and the detector positions relative to the beam and non-perfect collimation may have an influence. However, this has not been exhaustively investigated and other measurements have not shown this effect. Here, additional studies are needed. The maximum difference observed was slightly above 10\,\%. The majority of observables are in agreement with the estimated uncertainties of the dose measurement and the simulation. Although material assignment becomes more complicated with anthropomorphic phantoms, it was demonstrated that, for both simple scan parameters and more advanced scan protocols, equivalent source models and the MC simulation are well suited for verification of personalized CT dosimetry techniques. This was confirmed after using the technique at a different CT in a clinical environment at the St\"adtisches Klinikum hospital in Braunschweig. The equivalent source model of the Toshiba Aquilion ONE has been derived from measurements with the mobile measurement setup; the shape of the BT filters differs significantly from the BT of the Optima 660. The validation measurements of one rotation in axial mode yield differences between the simulation and the measurement of less than 10\,\%. Furthermore, multiple rotations in spiral mode were performed and the results obtained previously were confirmed. The major difference between the two scanners was related to large fluctuations of 8.5\,\% in the dose measurement when repeating single measurements. Since the different detector systems have been extensively tested and calibrated, the uncertainty of the air kerma measurement has been shown to be lower than 2\,\% ($k=1$) for the different detector systems. Hence, the fluctuation is very likely caused by the CT itself. The fluctuation of the tube output is evidently a disadvantage for any scan-specific or patient-specific dosimetry approach and leads to very high uncertainties. Additional studies are needed to investigate this effect and whether other clinical CT machines have comparable characteristics. The major sources of uncertainty for the simulation are related to the shape of the BT filter and the rotation angle of the X-ray tube when starting the irradiation. The lack of knowledge of the starting angle was identified to be one of the crucial challenges in order to achieve accurate simulation results. To overcome this problem, several simulations with different starting angles were performed. The results were averaged, which reduced the uncertainty. The effect of the geometry of the BT filter has been investigated by modifying the filter thickness with respect to the estimated uncertainties from the measurement of the total filtering. This effect is estimated to be less than 3\,\%, which is the same size as the uncertainty of the total filtration.

To finalize the complete procedure of personalized dosimetry, techniques for organ segmentation need to be developed and implemented as well as the correction for missing scattered radiation due to the shortness of the phantom. Additional verifications of the procedure are necessary using different CTs from other manufacturers (e.g. Siemens) and more advanced phantoms such as CIRS: Model 600 ``3D Sectional Torso Phantom'', as well as patient scan data for the final test. This will allow more realistic geometries and materials to be incorporated.

\section{Conclusion.}
\label{sec:Conclusion}
Viable, rapid procedures were developed that allow the post-CT scan dose to be measured and calculated at five positions inside anthropomorphic phantoms. Measured and calculated dose values inside the phantoms generally agreed within the evaluated uncertainties of less than 10\,\%. The procedures are applicable to any scanner type under clinical conditions. Results show that the procedures are well suited for verifying the applicability of personalized CT dosimetry based on post-scan Monte Carlo calculations. The procedures could become part of a potential acceptance test if personalized CT dosimetry is incorporated into future CT scanners.

\begin{acknowledgments}
The authors thank Reinulf B\"ottcher, Hartmut Drehsler, Marvin Wenzel and Lars Herrman (PTB) for supporting the measurements and mechanical construction of the mobile measurement equipment. 
Research funding: This project has received funding from the EMPIR programme 15HLT05, ``Metrology for multi-modality imaging of impaired tissue perfusion''. The EMPIR initiative is co-funded by the European Union's Horizon 2020 research and innovation programme and the EMPIR Participating States.
   
	

\end{acknowledgments}


\end{document}